\documentclass[12pt, letterpaper, top=3cm, bottom=3cm, right=2.5cm, left=2.5cm]{SimpleNotes}

\usepackage{natbib}
\usepackage{enumerate}
\usepackage{float}
\usepackage{dsfont}
\usepackage{algorithm}
\usepackage{algorithmicx}
\usepackage[titletoc,toc,title,page]{appendix}
\usepackage{tikz}
\usetikzlibrary{shapes,arrows}
\usepackage[font=small,labelfont=bf,justification=centering]{caption}
\usepackage{multirow}
\usepackage{ctable} 
\usepackage{booktabs}
\usepackage{tablefootnote}
\usepackage[nodisplayskipstretch]{setspace}
\usepackage{authblk}

\usepackage{bm} 
\usepackage{dsfont} 
\usepackage{amsmath} 
\usepackage{amssymb} 
\usepackage{xifthen} 

\newcommand{\rv}[1]{#1} 
\newcommand{\rve}[1]{\bm{#1}} 
\newcommand{\gp}{w} 
\newcommand{\mgp}{\bm{w}} 
\newcommand{\stp}[2][]{\left\lbrace#2\ifthenelse{\isempty{#1}}{}{:{#1}}\right\rbrace} 

\newcommand{\ve}[1]{\bm{#1}} 
\newcommand{\m}[1]{\bm{#1}} 

\newcommand{\tr}{\intercal}

\newcommand{\diag}[1]{\text{diag}(#1)}

\newcommand{\cross}[2][]{#2^\tr\ifthenelse{\isempty{#1}}{}{#1} #2}
\newcommand{\tcross}[2][]{#2\ifthenelse{\isempty{#1}}{}{#1} #2^\tr}
\newcommand{\mve}[1]{\text{vec}({#1})} 

\newcommand{\ind}[1]{\mathds{1}_{\left(#1\right)}} 

\newcommand{\real}{\mathbb{R}} 

\newcommand{\mean}[2][]{\mathds{E}\ifthenelse{\isempty{#1}}{}{_{#1}}\left[#2\right]}
\newcommand{\meanhat}[2][]{\hat{\mathds{E}}\ifthenelse{\isempty{#1}}{}{_{#1}}\left[#2\right]}
\newcommand{\var}[2][]{\mathds{V}\ifthenelse{\isempty{#1}}{}{_{#1}}\left[#2\right]}
\newcommand{\cov}[3][]{\text{Cov}\ifthenelse{\isempty{#1}}{}{_{#1}}\left[#2,#3\right]}
\newcommand{\cor}[3][]{\text{Cor}\ifthenelse{\isempty{#1}}{}{_{#1}}\left[#2,#3\right]}
\newcommand{\pr}[1]{\text{Pr}\left(#1\right)} 

\newcommand{\df}[1]{\mathcal{#1}} 

\newcommand{\degrees}{^{\circ}}
\newcommand{\eq}{\text{Equation }} 
\newcommand{\se}{\text{Section }} 
\newcommand{\fig}{\text{Figure }} 
\newcommand{\tab}{\text{Table }} 

\newcommand{\sm}{SM} 

\title{Mapping food insecurity in the Brazilian Amazon using a spatial item factor analysis model}

\author[1]{\normalsize Erick A. Chac\'{o}n-Montalv\'{a}n}
\author[2,3]{\normalsize Luke Parry}
\author[4]{\normalsize Emanuele Giorgi}
\author[5]{\normalsize Patricia Torres}
\author[6]{\normalsize Jesem D. Orellana}
\author[1]{\normalsize Paula Moraga}
\author[7]{\normalsize Benjamin M. Taylor}

\affil[1]{\footnotesize {Computer, Electrical and Mathematical Sciences and Engineering Division, King Abdullah University of Science and Technology (KAUST), Thuwal 23955-6900, Saudi Arabia}}
\affil[2]{\footnotesize {Instituto Amazônico de Agriculturas Familiares, Universidade Federal do Pará, Belém, Brazil}}
\affil[3]{\footnotesize {Lancaster Environment Centre, Lancaster University, United Kingdom.}}
\affil[4]{\footnotesize {Centre for Health Informatics, Computing, and Statistics (CHICAS), Lancaster Medical School, Lancaster University, United Kingdom.}}
\affil[5]{\footnotesize {World Resources Institute (WRI) Brasil, São Paulo, Brazil}}
\affil[6]{\footnotesize {Leonidas and Maria Deane Institute, Oswaldo Cruz Foundation, Manaus, Brazil}}
\affil[7]{\footnotesize {School of Mathematical Sciences, University College Cork, Ireland}}

\date{}

\begin{document}
\maketitle

\begin{abstract}[Continuous spatial variation, Factor analysis, Gaussian processes, Kriging, Model-based geostatistics, Structural equation modeling.]

Food insecurity, a latent construct defined as the lack of consistent access to sufficient
and nutritious food, is a pressing global issue with serious health and social justice
implications. Item factor analysis is commonly used to study such latent constructs, but
it typically assumes independence between sampling units. In the context of food
insecurity, this assumption is often unrealistic, as food access is linked to
socio-economic conditions and social relations that are spatially structured. To address
this, we propose a \textit{spatial item factor analysis} model that captures spatial
dependence, allowing us to predict latent factors at unsampled locations and identify food
insecurity hotspots. We develop a Bayesian sampling scheme for inference and illustrate
the explanatory strength of our model by analysing household perceptions of food
insecurity in Ipixuna, a remote river-dependent urban centre in the Brazilian Amazon. Our
approach is implemented in the R package \texttt{spifa}, with further details provided in
the Supplementary Material. This spatial extension offers policymakers and researchers a
stronger tool for understanding and addressing food insecurity to locate and prioritise
areas in greatest need. Our proposed methodology can be applied more widely to other
spatially structured latent constructs.

\end{abstract}


\section{Introduction}
\label{spifa_sec:introduction}

\subsection{Food insecurity and item response theory}
\label{ssub:fi_irt}

\textit{Food security} is a latent construct that characterizes a `situation that exists
when all people, at all times, have physical, social and economic access to sufficient,
safe and nutritious food that meets their dietary needs and food preferences for an active
and healthy life' \citep[pp 313]{fao2003}. The latent construct \textit{food insecurity}
describes the opposite situation, in which individual or household access to sufficient,
safe and nutritious food is unreliable or insufficient.
Understanding the level of
food insecurity in a region is crucial for designing appropriate public health and nutrition interventions.

Food insecurity has been traditionally studied using \textit{item response theory} (IRT),
which is a family of statistical models used to relate \textit{responses to items} to a
\textit{latent construct}. These models assume the latent construct (e.g. food insecurity)
or \textit{ability} is defined on a continuum. This allows us, for instance to score each
individual's level or ability; to identify which items have the greatest capacity to
\textit{discriminate} between individuals of differing abilities or levels (i.e. how well
each item identifies the trait of food insecurity); or to identify the \textit{difficulty}
or easiness associated to each item---more `difficult' items in this context would tend to
be endorsed by more food-insecure individuals, but less often by food-secure individuals
\citep{ayala2022theory}.

Item response theory is widely applied across diverse fields of research beyond food
insecurity analysis. In psychometrics, for example, it has been used to measure the theory
of mind ability \citep{black2019irt}, emotional intelligence \citep{Fiori2014}, and
self-esteem \citep{crowe2018selfesteem}. In health and medicine, it is used to determine
the health status of patients using self-reported outcomes
\citep{proust-lima2022modeling}, to measure individual scores of child developmental
status \citep{Drachler2007} and to assess achievement and evaluation of clinical
performance \citep{dechamplain2010primer}. In mental health research, it has been used to
study disorders like psychopathy \citep{tsang2018comparison}, alcohol use
\citep{saha2020performance} and depression \citep{stover2019state}. In e-learning, item
response theory has been used to develop personalized intelligent tutoring systems that
match learner ability and difficulty level \citep{pliakos2019integrating}. In computerized
adaptive testing, it is used in tests like GMAT, GRE or TOEFL to dynamically select the
most appropriate items for examinees according to individual abilities
\citep{jia2020design}. In marketing, it has been used to measure customer relationship
satisfaction \citep{peng2019multifacet} and to measure extreme response styles (ERS)
\citep{DeJong2008}. In criminology, it is applied to the analysis of the causes of crime
and deviance using self-reporting measures of delinquency \citep{thomas2019rationalizing}
and to measure self-control \citep{manapat2021psychometric}.

\subsection{Food insecurity in Ipixuna, Brazil}

\begin{figure}[!htb]
  \centering
  \includegraphics[width=\linewidth]{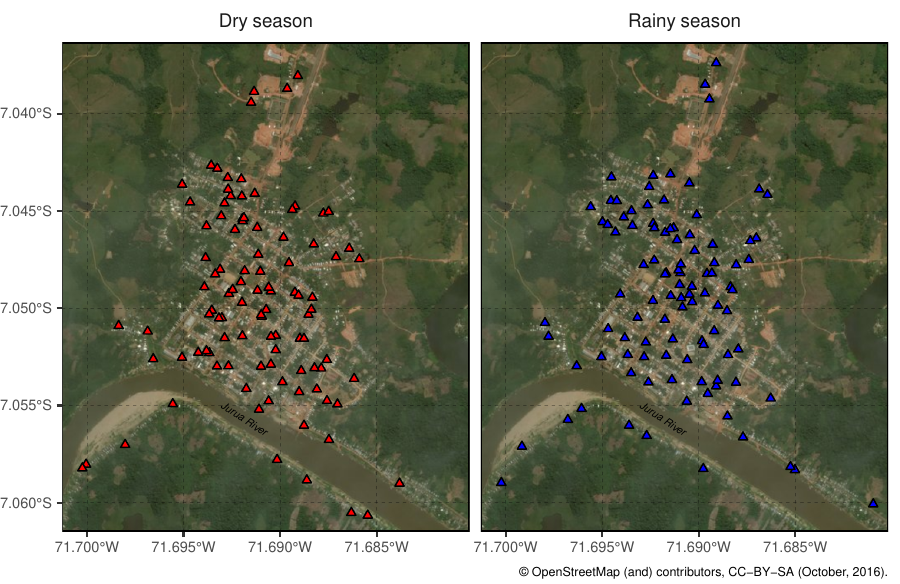}
  \caption{Spatial distribution of sampled households (triangles) in the urban area of
  Ipixuna for each season. The locations have been jittered to preserve respondent
  privacy.}
  \label{spifa_fig:ipixuna}
\end{figure}

Our motivating application concerns the assessment of household food insecurity which is
mediated through a family's ability to access food and also through the supply of food
potentially available. Both factors are relevant in the context of our study located in
Ipixuna, shown in \fig \ref{spifa_fig:ipixuna}, which is a `jungle town' located on the
banks of the River Juru\'a with a urban population of approximately twelve thousand
residents at the time of fieldwork \citep{ibge2010instituto,ibge2022instituto}; it is
unconnected to the Brazilian road network, and is several thousand kilometers of upstream
boat travel from the Amazonas State capital, Manaus. Being extremely remote and highly
river-dependent, Ipixuna exhibits high social vulnerability to hydro-climatic shocks such
as floods and droughts, which pose a serious risk of harm to the local population
\citep{parry2018social}. This vulnerability is due to a combination of high social
sensitivity (e.g. low levels of formal education, many households lack access to clean
piped water and internal toilets), low adaptive capacity (e.g. relatively poor public
services, proxied by educational delays and inadequate antenatal care), and high prices of
imported food stuffs (e.g. frozen chicken) \citep{parry2018social}.


In these remote and roadless urban centres, accessible only by boat or plane, food
insecurity is partly affected by seasonal variation in river levels. All river-dependent
towns and cities in Amazonas State experience strong seasonal flood pulses (reflecting
highly seasonal rainfall patterns, including in locations far upstream in neighbouring
Amazonian countries; \citet{chacon-montalvan2021rainfall}) which cause dramatic changes in
river-levels and hence the relative ease or difficult of boats navigating between urban
centres. However, Ipixuna's location far up the River Juruá means that this town tends to
experience significant problems in boat-accessibility even in `normal' dry seasons. During
periods of extreme drought such as in 2023 \citep{santosdelima2024severe}, Ipixuna and
other urban centres like it became inaccessible to larger boats for several months.
Conversely, the annual high-water flood season carries risks of rising above the limits of
the stilted houses at the urban periphery and even flood the urban centre. In extreme
cases (such as early 2017) houses in Ipixuna's poorer riverine neighbourhoods flood and
become uninhabitable, with associated
risks for health (e.g. waterborne diseases) and income (i.e. disrupted livelihoods). These
climate-health risks explain why prenatal exposure to periods of extremely deficient or
intense rainfall in Amazonas lead to higher odds of pre-term birth and restricted
intra-uterine growth \citep{chacon-montalvan2021rainfall}. But there are other factors at
play too: community, governmental and non-governmental support can bolster a family's
access to food resources in difficult times \citep{garrett1999,battersby2011}, and social
inequities shape household's capacities for coping with environmental shocks. As is the
case with cities in the Global North, neighbourhoods with certain characteristics tend to
cluster together. In Ipixuna, the poorest neighbourhoods are at the urban margins, home to
many rural-urban migrants. The arrival of such migrants to Amazonian towns from the 1980s
onwards followed the collapse of the rubber economy and rural people's desires to access
basic services, especially education \citep{Parry2010}.
It is therefore unsurprising that Ipixuna has rapidly urbanized; its urban population
expanded by $65\%$ over the decade to 2010 and by another $58\%$ by 2020
\citep{ibge2010instituto, ibge2022instituto}.

We measured food insecurity items in August 2015 (low-water dry season) and March 2016
(high-water rainy season) with the male or female heads of 100 randomly sampled households per season (see \fig
\ref{spifa_fig:ipixuna}). We measured perceptions of household food insecurity using
a questionnaire that we modified from the Brazilian Household Food Insecurity Scale
(EBIA) \citep[see][Supplementary Materials]{rivero2022urban}. The EBIA was developed and
validated in Brazil in 2003, and built on the Household Food Security Survey Module
(HFSSM) from the US Department of Agriculture \citep{perez-escamilla2004adapted}. The EBIA
has similarities with the widely-used Food Insecurity Access Scale (HFIAS), which also
originated from the HFSSM, and was designed by USAID to be adapted for diverse cultural
contexts \citep{coates2007household}. In our study, we asked about experiences during the
previous 30 days in order to obtain season-specific food insecurity measures, consistent
with our sampling of peak rainy and dry seasons.
The full questionnaire, which contains 18 items related to food insecurity, is available
in the Supplementary Material (\sm); see \citet[\se S3]{chacon-montalvan2025supplement}.
Items in Section A referred to household as
a whole, those in Section B referred to adults only, Section C concerned children (those
under the age of 18), and Section D included items related to the regional context of our
study. The regionally-specific questions in section D were designed to measure similar
aspects as contained in the general scale but were adapted to reflect common coping
strategies employed in this locality. It is common for observed data from household
surveys to include missing data. This can be due to human error in recording information,
but more importantly in our application, due to the presence of items in Section C that
only apply to households that have children.

We revisited Ipixuna in May 2017 in order to share our provisional food insecurity results
with diverse stakeholders, and use these findings (including maps and neighbourhood-scale
summaries of food insecurity challenges) as a starting point for discussing food
insecurity in this municipality in relation to key themes which included social
marginalization, socio-spatial inequities, governance, seasonality (particularly in
relation to local river levels, but also rainfall) and extreme hydro-climatic events. Our
multi-disciplinary team had expertise in geography, epidemiology and nutrition, rural
livelihoods, and hydrology, and citizen participants in our pan-Amazon `Amazonas Citizens
Network' (\url{http://wp.lancs.ac.uk/rede-cidada-am/eventos/}). Activities over our 12-day visit included a three-day workshop in the town (with representatives from diverse municipal and state-level institutions including Civil Defence, churches, etc), field-visits to neighbourhoods (especially more vulnerable peri-urban areas), and semi-structured interviews with key respondents (e.g. with the local head of the Amazonas Institute for Agricultural Development and Sustainable Forestry (IDAM)).

In our application, we are interested in estimating the difficulty and discrimination
parameters in order to understand the relationship between the underlying latent factors
with the items, using all the available data, including those households with missing
data. We also aim to predict the latent factors not only in places where the observations
were taken, but also in locations where we have no observations. Collecting our data was
costly, difficult, and time-consuming, thus our method for predicting food insecurity at
new locations is an important step for identifying particularly vulnerable areas that
could be targeted for intervention. Since our method can also be used to map and predict
the different dimensions of food insecurity, this information could be used to tailor
specific interventions to specific regions.

\subsection{Limitation of item response theory and item factor analysis}

One of the main limitations of classical IRT models is that they assume that the latent construct is unidimensional: this assumption may not be adequate for more complex latent constructs. For example, the items developed to study food insecurity capture a number of different concepts including: (i) the perception of reduction in the quality or quantity of food, (ii) an actual reduction in quality of food,  (iii) an actual reduction in quantity of food, and (iv) a reduction in the quantity or quality of food for children in the household. Hence, the construct food insecurity has more than one dimension, and might also depend on characteristics of the population under study, \citet{froelich2002dimensionality} for example found a further dimension associated with the protection of children from hunger.

In this context, where unidimensional models are not appropriate, researchers have
developed \textit{Multidimensional Item Response Theory} (MIRT) or \textit{Item Factor
Analysis} (IFA), both approaches being conceptually similar \citep{Bock1988,
Chalmers2012a}. These models extend the concept of standard multivariate factor analysis
so it can be applied to binary or ordinal data and allows us to study the interaction
between multiple items and a multi-dimensional latent construct.
Although IFA overcomes the limitation of unidimensionality, it has other constraints that we seek to address in this paper.

Firstly, IFA assumes the latent construct of a particular subject to be independent of any
other subject. In our subsequent example of food insecurity, this seems inadequate given
that households near to each other are more likely to have similar socio-economic
conditions and environmental exposures, share food between households, and thus a similar
risk of food insecurity. This observation also applies to the analysis of latent
constructs in other disciplines where spatial correlation is naturally expected, an
example would be socio-economic status itself. Connected to this, an IFA model incorporating spatial random effects would allow us to map the latent factors at unobserved locations, which can be (and is in our case) of scientific interest. With respect to our own and other similar application(s), a complete map of the latent factors over the area under study will improve our understanding of the construct and help to better inform the decision-making process.

Secondly, IFA only relates items to the latent construct, but not to possible covariates that could help explain why certain individuals might have particularly high or low values of the latent construct. For example, our previous research in this area suggests that accessibility and political economic factors play an important role in determining social vulnerability in Amazonian towns and cities \citep{parry2018social}.
In our case, therefore, understanding the relationship between the items, the latent construct and the covariates is highly desirable.

Finally, traditional methods often require removing profiles with missing data or applying
simple imputation methods that do not account for the uncertainty associated with this
imputation. Therefore, we need an approach that allows us to use the full data set while
considering the uncertainty associated with the missing data.

\subsection{Spatial item factor analysis}

The above summarises our motivation for developing an extension to IFA which we here
denominate \textit{spatial item factor analysis} (SPIFA). Some extensions of factor analysis to
the spatial domain have been undertaken, for example \citet{Wang2003, Frichot2012,
thorson2015spatial}; however, they are not adequate for our study because the first
considers only one latent factor; the second, includes the spatial structure only on the
residual term; while the last, considers a spatial structure only on the latent factor. In
comparison, our hierarchical framework allows the latent construct to be split into
multiple latent factors, the number and composition of which are determined by initial
exploratory analyses or expert knowledge. These latent factors are explained by observed covariates,
spatially-correlated random effects and non-spatial multivariate random effects. The
relationship between the latent factors and the item responses, in the case of binary
outcomes, is mediated through a set of auxiliary variables which handle the conversion
between continuous to discrete data forms.

Additionally, our model accounts for the presence of missing data and the uncertainty
associated when imputing values by treating them as latent variables and sampling them
when required \citep[see][\se S2]{chacon-montalvan2025supplement}. Although our model is
motivated by our interest in mapping food insecurity, it can be applied more broadly in
settings where IFA is used, particularly when the latent construct is expected to exhibit
spatial structure---something that can also be explored through preliminary analysis.

\subsection{Structure}

Details of our proposed SPIFA model are presented in \se
\ref{spifa_sec:spatial_item_factor_analysis}, and the model is implemented in our
\texttt{R} package \texttt{spifa}. We then present the application of the model for
predicting food insecurity in the urban town of Ipixuna, Brazil, in \se
\ref{spifa_sec:predicting_food_insecurity_on_the_brazilian_amazonia}. Finally, the paper
concludes with a discussion of the advantages, limitations, and possible extensions of our
model in \se \ref{spifa_sec:discussion}, along with the main findings from the analysis of
food insecurity in Ipixuna.


\section{Spatial Item Factor Analysis}
\label{spifa_sec:spatial_item_factor_analysis}

In this section, we develop the modelling framework for SPIFA. We first introduce
classical IFA in \se \ref{spifa_sub:introduction_ifa} and \se \ref{spifa_sub:eifa_cifa},
and then present our new methods in \se \ref{spifa_sub:extension_to_the_spatial_domain}.
Solutions to identifiability issues are discussed in \se
\ref{spifa_sub:identifiability_and_restrictions}, while further flexibility in the
multivariate spatial structure is explained in \se
\ref{spifa_sub:allowing_further_flexibility_on_the_multivariate_spatial_structure}. We
then provide details about the identifiable auxiliary variables in \se
\ref{spifa_sub:auxiliary_variables_identifiable_spifa} and their matrix form in \se
\ref{spifa_sub:matrix_form_of_the_auxiliary_veatiables}. We conclude with the
specification of the likelihood function in \se \ref{spifa_sub:likelihood_function}, and
information about our \texttt{spifa} R package in \se \ref{spifa_sec:r_package}.

\subsection{Item factor analysis}
\label{spifa_sub:introduction_ifa}

IFA can be seen as an extension of factor analysis for binary or ordinal data. In this
article, we focus on binary outcomes and discuss possible extensions of the proposed
framework to mixed data types---continuous, binary, and ordinal---in the \sm{} \citep[\se S5]{chacon-montalvan2025supplement}.

We begin by considering the response variable $\rv{Y}_{ij}$ for item $j=1,2,\dots,q$ from
subject $i=1,2,\dots,n$ as a binarization around zero of a continuous but unobservable
process $\rv{Z}_{ij}$, explained by $m$ latent factors $\theta_{ik}$, also called latent
abilities,
\begin{align}
  \label{spifa_eq:ifa_aux_var}
  \rv{Z}_{ij} & = c_j + \sum_{k = 1}^m a_{jk}\theta_{ki} + \epsilon_{ij},
\end{align}
where $\epsilon_{ij} \sim \df{N}(0, 1)$ and $\{c_j\}$ are intercept parameters that
reflect difficulty of items. High positive (negative) values for $c_j$ increase (reduce)
the probability of endorsing $j$-th item, which is why they are also referred to as
\emph{easiness} parameters \citep{Chalmers2015}. The elements of the slope vector
$\ve{a}_j = (a_{j1}, \dots, a_{jm})^\tr$, commonly called \emph{discrimination}
parameters, indicate how well the $j$-th item can discriminate the $k$-th ability among
the subjects under study. If $a_{jk}=0$, then the $k$-th latent factor does not explain
the variability of the $j$-th response item, meaning that this item does not help to
discriminate the $k$-th latent ability among the subjects. In our work, we use this same
parameterisation with intercepts and slopes. Further details on this model, including
inference via the expectation-maximization algorithm, can be found in \citet{Bock1988}.

Beyond estimating the easiness and discrimination parameters, another goal is to make
inferences for the latent factor $\ve{\theta}_i = (\theta_{1i}, \dots, \theta_{mi})^\tr$.
This allows us to differentiate individuals with high or low levels of the construct under
study. A practical application is in the area of \emph{ideal point estimates}, where the
objective is to estimate the ideological position of a legislator to predict whether they
will vote in favour of a particular motion \citep{Bafumi2005}.

\subsection{Exploratory and confirmatory item factor analysis}
\label{spifa_sub:eifa_cifa}

The IFA model defined by \eq \eqref{spifa_eq:ifa_aux_var} is not identifiable due to
various types of aliasing (see \se \ref{spifa_sub:identifiability_and_restrictions}). To
achieve identifiability, restrictions must be imposed on certain parameters and the latent
factors, which can be done in ways that lead to either exploratory or confirmatory
analyses.

An exploratory item factor analysis (EIFA) is obtained when restrictions are imposed
solely to enable inference; in this case, the restrictions are not tied to the specific
construct or dataset under study. The usual restrictions are that the latent factors
follow a standard multivariate normal distribution, $\ve{\theta}_i \sim \df{N}(\ve{0},
\m{I})$, and that the matrix of discrimination parameters, $\m{A} = (\ve{a}_1, \dots,
\ve{a}_q)^\intercal$, is lower triangular \citep{Cai2010}. Estimates can then be rotated
according to the researcher’s preference.

By contrast, a confirmatory item factor analysis (CIFA) applies restrictions in a
semi-formal manner: the researcher draws on expert knowledge to define a plausible
structure for the latent construct. The main restrictions are that the latent factors
follow a zero-mean multivariate normal distribution with unit variances, $\rve{\theta}_i
\sim \df{N}(\ve{0}, \m{\Sigma}_\theta)$, where $\diag{\m{\Sigma}_\theta} = \ve{1}$, and
that the researcher imposes at least $m(m-1)/2$ zero constraints on $\m{A}$
\citep{Cai2010a}. In CIFA, the restrictions are tailored to the specific study and its
context, whereas in EIFA they are general and not problem-specific. Where there is no
consensus among experts about the structure, researchers can still compare candidate
models using measures of fit (e.g., WAIC or DIC for Bayesian analysis).

\subsection{Extension to the spatial domain}
\label{spifa_sub:extension_to_the_spatial_domain}

The extension of IFA to the spatial domain can be achieved by including a spatial process
in \eq \eqref{spifa_eq:ifa_aux_var}. \citet{Frichot2012}, for example, proposed a model
that includes a spatially correlated error term $\epsilon_{ij}$ to correct the principal
components by modeling the residual spatial variation. Similarly to \citet{Wang2003} and
\citet{thorson2015spatial}, our proposed SPIFA method allows the latent factors
$\theta_{ki}$ to be spatially correlated because the nature of the particular construct
under study suggests they should be treated in this way. For example, spatial patterns in
food insecurity scores are expected across a municipality due to their association with
socio-economic and environmental variables. \citet{Wang2003} represent the latent factor
as a uni-dimensional spatial process, while \citet{thorson2015spatial} represent it as a
multi-dimensional spatial process with a lower-triangular structure for the discrimination
parameters and employ integrated nested Laplace approximation for inference. In contrast,
our model allows the multi-dimensional latent factor to be explained by predictors, a
multivariate spatial process, and a multivariate non-spatial effect. The discrimination
parameters in our model are flexible and can be defined by the researcher based on the
area of analysis. Additionally, our approach employs full Bayesian inference using MCMC
methods. The details of our model are explained below.

We model the binary response variables as discrete-state stochastic processes
$\{\rv{Y}_j(s): s \in D\}$ where $D\subset\real^2$ and $\rv{Y}_j(s)$ denotes the
response to item $j$ at spatial location $s$. The response variables take the values $0$ or
$1$ according to the value of an underlying auxiliary spatial process
$\{\rv{Z}_j(s): s \in D \}$:
\begin{align}
  \label{spifa_eq:def_response}
  \rv{Y}_{j}(s) & =
  \left\lbrace
  \begin{array}[2]{cc}
    1 & \text{if} ~ \rv{Z}_{j}(s) > 0, \\
    0 & \text{otherwise}.
  \end{array}
  \right.
\end{align}
Note that $\rv{Y}_j(s)$ is deterministic conditional on $\rv{Z}_j(s)$, which is defined as
follows:
\begin{align}
  \label{spifa_eq:def_auxiliary}
  \rv{Z}_{j}(s) & =
  c_j + \ve{a}_j^\intercal\ve{\theta}(s) + \epsilon_{j}(s),
  ~~ \epsilon_{j}(s) \sim \df{N}(0, 1),
\end{align}
where $c_j$ and $\ve{a}_j$ are the easiness and discrimination parameters, respectively.
The $m$-dimensional latent factor $\ve{\theta}(s)$ is defined as a function of covariates,
a multivariate spatial process, and a non-spatially correlated error term, as specified in
\eq \eqref{spifa_eq:def_abilities}. The latent factor is the only source of spatial
correlation in $\rv{Z}_j(s)$ and hence in $\rv{Y}_j(s)$: if the spatial variation is
removed from $\rve{\theta}(s)$, then the model resembles a standard IFA.

Since one of our aims is to predict the latent factors at unobserved locations, part of
the variability of $\rve{\theta}(s)$ is explained by a set of spatial covariates
$\ve{x}(s) = (x_1(s), \dots, x_p(s))^\tr$ that are preferably also available at unobserved
sites. This allows the model to help us understand why certain locations have high or low
scores---similar in spirit to \textit{multiple indicators, multiple causes models}
\citep{Tekwe2014}. We also include a spatial $m$-dimensional process $\{\mgp(s): s \in
D\}$ to capture the remaining spatial structure in $\{\rve{\theta}(s)\}$ after accounting
for the covariate effects; and a non-spatial $m$-dimensional term $\ve{v}(s) \sim
\df{N}(\ve{0}, \m{\Sigma}_v)$ to represent additional variation. The latent factor is thus
defined as:
\begin{equation}
  \label{spifa_eq:def_abilities}
  \ve{\theta}(s) = \ve{\eta}(s) + \mgp(s) + \ve{v}(s),
\end{equation}
where $\ve{\eta}(s) = \m{B}^\tr\ve{x}(s)$ represents the covariate effects, with $\m{B}$ a
$p \times m$ matrix of slopes associating the covariates $\ve{x}(s)$ with the latent
factors $\rve{\theta}(s)$. Note that the covariates are assumed to be standardised; see
Section \ref{spifa_sub:identifiability_and_restrictions} for further details.

The $m$-dimensional spatial process $\{\mgp(s)\}$ is defined as a collection of $m$
zero-mean, independent, stationary and isotropic Gaussian processes with variance
$\sigma^2_k$ and correlation function $\cor{\gp_k(s)}{\gp_k(s')} = \rho_k(u)$, where $u$
is the distance between two spatial locations $s$ and $s'$,
\begin{equation}
  \label{spifa_eq:gp}
  \{\gp_k(s)\} \sim \text{GP}(0, \sigma^2_k, \rho_k(u)), ~~~~ k = 1, \dots, m.
\end{equation}
While this definition might appear restrictive, the independence assumption for these
spatial processes is appropriate when the latent factors $\rv{\theta}_k(s)$ are
independent. This can also be extended to cases where the latent factors are correlated,
as described in Section
\ref{spifa_sub:allowing_further_flexibility_on_the_multivariate_spatial_structure}, where
the multivariate notation $\{\mgp(s)\}$ is used to explore possible extensions to its structure.
Consequently, we consider it a multivariate Gaussian process (MGP).

\eq \eqref{spifa_eq:def_abilities} has the same structure as a multivariate geostatistical
model. However, in our case, the dependent variable, $\rve{\theta}(s)$, is a
low-dimensional latent process instead of a high-dimensional observed process as in
\citet{Gelfand2004}. A similar structure including fixed and random effects is also
discussed in \citet{Chalmers2015}, but the author does not attempt to model unexplained
spatial variation. In addition, the author mainly focuses on including covariates at the
item level, whereas our emphasis is on the inclusion of spatial covariates which will then allow us to make predictions at unobserved locations.

Substituting the structure of the latent factors $\ve{\theta}(s)$ into \eq \eqref{spifa_eq:def_auxiliary} results in
\begin{align}
  \label{spifa_eq:def_auxiliary_marginal}
  \rv{Z}_{j}(s) & = c_j + \ve{a}_j^\tr[\m{B}^\tr\ve{x}(s) + \mgp(s) + \ve{v}(s)] + \epsilon_{j}(s).
\end{align}
We note that if $\ve{a}_j$ were known, then \eq \eqref{spifa_eq:def_auxiliary_marginal}
would be a multivariate geostatistical model, allowing for classical inference methods.
However, the main challenges in performing inference for our proposed model come from the
inclusion of the interaction between the latent variables with the (unknown) slopes,
$\ve{a}_j$, as this interaction introduces identifiability issues, as discussed in \se
\ref{spifa_sub:identifiability_and_restrictions}.

\begin{figure}[!ht]
  \centering
  \begin{tikzpicture}[>=latex,line join=bevel,scale=0.8]
\begin{scope}
  \pgfsetstrokecolor{black}
  \definecolor{strokecol}{rgb}{1.0,1.0,1.0};
  \pgfsetstrokecolor{strokecol}
  \definecolor{fillcol}{rgb}{1.0,1.0,1.0};
  \pgfsetfillcolor{fillcol}
\end{scope}
\begin{scope}
  \pgfsetstrokecolor{black}
  \definecolor{strokecol}{rgb}{1.0,1.0,1.0};
  \pgfsetstrokecolor{strokecol}
  \definecolor{fillcol}{rgb}{1.0,1.0,1.0};
  \pgfsetfillcolor{fillcol}
\end{scope}
\begin{scope}
  \pgfsetstrokecolor{black}
  \definecolor{strokecol}{rgb}{1.0,1.0,1.0};
  \pgfsetstrokecolor{strokecol}
  \definecolor{fillcol}{rgb}{1.0,1.0,1.0};
  \pgfsetfillcolor{fillcol}
\end{scope}
  \node (f1) at (57.0bp,120.0bp) [draw,ellipse] {$\theta_1$};
  \node (f2) at (189.0bp,120.0bp) [draw,ellipse] {$\theta_2$};
  \node (f3) at (321.0bp,120.0bp) [draw,ellipse] {$\theta_3$};
  \node (f4) at (461.0bp,120.0bp) [draw,ellipse] {$\theta_4$};
  \node (y9) at (365.0bp,7.5bp) [draw,rectangle] {$Y_9$};
  \node (y8) at (321.0bp,7.5bp) [draw,rectangle] {$Y_8$};
  \node (x3) at (238.0bp,228.5bp) [draw,rectangle] {$x_{3}$};
  \node (y1) at (13.0bp,7.5bp) [draw,rectangle] {$Y_1$};
  \node (y3) at (101.0bp,7.5bp) [draw,rectangle] {$Y_3$};
  \node (y2) at (57.0bp,7.5bp) [draw,rectangle] {$Y_2$};
  \node (y5) at (189.0bp,7.5bp) [draw,rectangle] {$Y_5$};
  \node (y4) at (145.0bp,7.5bp) [draw,rectangle] {$Y_4$};
  \node (y7) at (277.0bp,7.5bp) [draw,rectangle] {$Y_7$};
  \node (y6) at (233.0bp,7.5bp) [draw,rectangle] {$Y_6$};
  \node (x1) at (168.0bp,228.5bp) [draw,rectangle] {$x_{1}$};
  \node (x2) at (203.0bp,228.5bp) [draw,rectangle] {$x_{2}$};
  \node (x4) at (273.0bp,228.5bp) [draw,rectangle] {$x_{4}$};
  \node (x5) at (308.0bp,228.5bp) [draw,rectangle] {$x_{5}$};
  \node (g4) at (418.0bp,120.0bp) [draw,fill=black!20!white,ellipse] {$w_4$};
  \node (g3) at (278.0bp,120.0bp) [draw,fill=black!20!white,ellipse] {$w_3$};
  \node (g2) at (146.0bp,120.0bp) [draw,fill=black!20!white,ellipse] {$w_2$};
  \node (g1) at (14.0bp,120.0bp) [draw,fill=black!20!white,ellipse] {$w_1$};
  \node (z8) at (321.0bp,62.0bp) [draw,ellipse] {$Z_8$};
  \node (z9) at (365.0bp,62.0bp) [draw,ellipse] {$Z_9$};
  \node (z4) at (145.0bp,62.0bp) [draw,ellipse] {$Z_4$};
  \node (z5) at (189.0bp,62.0bp) [draw,ellipse] {$Z_5$};
  \node (z6) at (233.0bp,62.0bp) [draw,ellipse] {$Z_6$};
  \node (z7) at (277.0bp,62.0bp) [draw,ellipse] {$Z_7$};
  \node (x6) at (343.0bp,228.5bp) [draw,rectangle] {$x_{6}$};
  \node (z1) at (13.0bp,62.0bp) [draw,ellipse] {$Z_1$};
  \node (z2) at (57.0bp,62.0bp) [draw,ellipse] {$Z_2$};
  \node (z3) at (101.0bp,62.0bp) [draw,ellipse] {$Z_3$};
  \node (eta4) at (461.0bp,176.5bp) [draw,ellipse] {$\eta_4$};
  \node (eta1) at (57.0bp,176.5bp) [draw,ellipse] {$\eta_1$};
  \node (eta2) at (189.0bp,176.5bp) [draw,ellipse] {$\eta_2$};
  \node (eta3) at (321.0bp,176.5bp) [draw,ellipse] {$\eta_3$};
  \node (z12) at (510.0bp,62.0bp) [draw,ellipse] {$Z_{12}$};
  \node (z10) at (412.0bp,62.0bp) [draw,ellipse] {$Z_{10}$};
  \node (z11) at (461.0bp,62.0bp) [draw,ellipse] {$Z_{11}$};
  \node (y11) at (461.0bp,7.5bp) [draw,rectangle] {$Y_{11}$};
  \node (y10) at (412.0bp,7.5bp) [draw,rectangle] {$Y_{10}$};
  \node (y12) at (510.0bp,7.5bp) [draw,rectangle] {$Y_{12}$};
  \draw [->] (x6) ..controls (331.35bp,223.67bp) and (328.05bp,222.72bp)  .. (325.0bp,222.0bp) .. controls (234.41bp,200.74bp) and (124.29bp,185.82bp)  .. (eta1);
  \draw [->] (x4) ..controls (284.29bp,223.73bp) and (287.26bp,222.8bp)  .. (290.0bp,222.0bp) .. controls (343.74bp,206.33bp) and (407.76bp,190.39bp)  .. (eta4);
  \draw [<->] (f1) ..controls (91.753bp,136.72bp) and (113.14bp,145.34bp)  .. (133.0bp,149.0bp) .. controls (202.06bp,161.73bp) and (225.9bp,175.32bp)  .. (291.0bp,149.0bp) .. controls (297.28bp,146.46bp) and (302.98bp,141.91bp)  .. (f3);
  \draw [->] (x4) ..controls (285.5bp,214.48bp) and (297.81bp,201.65bp)  .. (eta3);
  \draw [->] (x5) ..controls (311.12bp,215.5bp) and (313.82bp,205.13bp)  .. (eta3);
  \draw [->] (f4) ..controls (461.0bp,101.91bp) and (461.0bp,92.171bp)  .. (z11);
  \draw [->] (eta2) ..controls (189.0bp,160.09bp) and (189.0bp,150.09bp)  .. (f2);
  \draw [->] (x2) ..controls (171.47bp,216.7bp) and (109.35bp,195.43bp)  .. (eta1);
  \draw [->] (z12) ..controls (510.0bp,43.667bp) and (510.0bp,33.805bp)  .. (y12);
  \draw [->] (x4) ..controls (261.32bp,223.78bp) and (258.03bp,222.8bp)  .. (255.0bp,222.0bp) .. controls (191.11bp,205.08bp) and (114.53bp,189.06bp)  .. (eta1);
  \draw [->] (x1) ..controls (141.4bp,215.52bp) and (100.17bp,196.95bp)  .. (eta1);
  \draw [->] (f2) ..controls (176.08bp,102.56bp) and (166.07bp,89.815bp)  .. (z4);
  \draw [->] (x5) ..controls (340.39bp,216.91bp) and (406.91bp,195.18bp)  .. (eta4);
  \draw [->] (x6) ..controls (310.39bp,216.91bp) and (243.44bp,195.18bp)  .. (eta2);
  \draw [->] (z10) ..controls (412.0bp,43.667bp) and (412.0bp,33.805bp)  .. (y10);
  \draw [->] (f1) ..controls (69.917bp,102.56bp) and (79.931bp,89.815bp)  .. (z3);
  \draw [<->] (f1) ..controls (95.577bp,144.58bp) and (130.42bp,160.56bp)  .. (159.0bp,149.0bp) .. controls (165.28bp,146.46bp) and (170.98bp,141.91bp)  .. (f2);
  \draw [->] (z6) ..controls (233.0bp,43.667bp) and (233.0bp,33.805bp)  .. (y6);
  \draw [->] (g2) ..controls (161.76bp,120.0bp) and (164.42bp,120.0bp)  .. (f2);
  \draw [->] (z4) ..controls (145.0bp,43.667bp) and (145.0bp,33.805bp)  .. (y4);
  \draw [->] (x2) ..controls (199.62bp,215.42bp) and (196.67bp,204.88bp)  .. (eta2);
  \draw [->] (g1) ..controls (29.759bp,120.0bp) and (32.416bp,120.0bp)  .. (f1);
  \draw [->] (x3) ..controls (259.92bp,214.3bp) and (285.96bp,198.61bp)  .. (eta3);
  \draw [->] (z11) ..controls (461.0bp,43.667bp) and (461.0bp,33.805bp)  .. (y11);
  \draw [->] (f4) ..controls (475.46bp,102.48bp) and (486.76bp,89.56bp)  .. (z12);
  \draw [->] (z5) ..controls (189.0bp,43.667bp) and (189.0bp,33.805bp)  .. (y5);
  \draw [->] (z7) ..controls (277.0bp,43.667bp) and (277.0bp,33.805bp)  .. (y7);
  \draw [->] (x6) ..controls (337.63bp,215.28bp) and (332.85bp,204.42bp)  .. (eta3);
  \draw [->] (f1) ..controls (57.0bp,101.91bp) and (57.0bp,92.171bp)  .. (z2);
  \draw [->] (x4) ..controls (250.75bp,214.25bp) and (224.19bp,198.45bp)  .. (eta2);
  \draw [->] (x1) ..controls (179.25bp,223.57bp) and (182.23bp,222.67bp)  .. (185.0bp,222.0bp) .. controls (278.5bp,199.48bp) and (392.62bp,185.16bp)  .. (eta4);
  \draw [->] (x5) ..controls (296.34bp,223.72bp) and (293.05bp,222.75bp)  .. (290.0bp,222.0bp) .. controls (212.73bp,202.88bp) and (119.32bp,187.26bp)  .. (eta1);
  \draw [->] (z1) ..controls (13.0bp,43.667bp) and (13.0bp,33.805bp)  .. (y1);
  \draw [->] (x2) ..controls (230.81bp,215.72bp) and (275.58bp,196.75bp)  .. (eta3);
  \draw [->] (x3) ..controls (225.24bp,214.48bp) and (212.67bp,201.65bp)  .. (eta2);
  \draw [->] (z3) ..controls (101.0bp,43.667bp) and (101.0bp,33.805bp)  .. (y3);
  \draw [->] (f3) ..controls (333.92bp,102.56bp) and (343.93bp,89.815bp)  .. (z9);
  \draw [->] (x1) ..controls (200.39bp,216.91bp) and (266.91bp,195.18bp)  .. (eta3);
  \draw [<->] (f2) ..controls (227.58bp,144.58bp) and (262.42bp,160.56bp)  .. (291.0bp,149.0bp) .. controls (297.28bp,146.46bp) and (302.98bp,141.91bp)  .. (f3);
  \draw [<->] (f1) ..controls (91.753bp,136.72bp) and (113.14bp,145.34bp)  .. (133.0bp,149.0bp) .. controls (198.13bp,161.0bp) and (369.61bp,173.82bp)  .. (431.0bp,149.0bp) .. controls (437.28bp,146.46bp) and (442.98bp,141.91bp)  .. (f4);
  \draw [->] (z9) ..controls (365.0bp,43.667bp) and (365.0bp,33.805bp)  .. (y9);
  \draw [->] (x6) ..controls (370.81bp,215.72bp) and (415.58bp,196.75bp)  .. (eta4);
  \draw [->] (f1) ..controls (44.083bp,102.56bp) and (34.069bp,89.815bp)  .. (z1);
  \draw [->] (g3) ..controls (293.76bp,120.0bp) and (296.42bp,120.0bp)  .. (f3);
  \draw [->] (z8) ..controls (321.0bp,43.667bp) and (321.0bp,33.805bp)  .. (y8);
  \draw [->] (f2) ..controls (201.92bp,102.56bp) and (211.93bp,89.815bp)  .. (z6);
  \draw [->] (z2) ..controls (57.0bp,43.667bp) and (57.0bp,33.805bp)  .. (y2);
  \draw [->] (eta3) ..controls (321.0bp,160.09bp) and (321.0bp,150.09bp)  .. (f3);
  \draw [->] (x5) ..controls (280.43bp,215.92bp) and (234.32bp,196.54bp)  .. (eta2);
  \draw [->] (eta4) ..controls (461.0bp,160.09bp) and (461.0bp,150.09bp)  .. (f4);
  \draw [->] (x3) ..controls (226.3bp,223.86bp) and (223.01bp,222.87bp)  .. (220.0bp,222.0bp) .. controls (169.05bp,207.27bp) and (108.77bp,191.15bp)  .. (eta1);
  \draw [->] (x3) ..controls (249.27bp,223.66bp) and (252.24bp,222.74bp)  .. (255.0bp,222.0bp) .. controls (321.6bp,204.08bp) and (401.81bp,188.42bp)  .. (eta4);
  \draw [->] (f3) ..controls (321.0bp,101.91bp) and (321.0bp,92.171bp)  .. (z8);
  \draw [->] (f4) ..controls (446.54bp,102.48bp) and (435.24bp,89.56bp)  .. (z10);
  \draw [->] (x1) ..controls (173.13bp,215.28bp) and (177.69bp,204.42bp)  .. (eta2);
  \draw [->] (f3) ..controls (308.08bp,102.56bp) and (298.07bp,89.815bp)  .. (z7);
  \draw [->] (f2) ..controls (189.0bp,101.91bp) and (189.0bp,92.171bp)  .. (z5);
  \draw [->] (g4) ..controls (433.76bp,120.0bp) and (436.42bp,120.0bp)  .. (f4);
  \draw [->] (eta1) ..controls (57.0bp,160.09bp) and (57.0bp,150.09bp)  .. (f1);
  \draw [->] (x2) ..controls (214.26bp,223.61bp) and (217.23bp,222.7bp)  .. (220.0bp,222.0bp) .. controls (300.27bp,201.7bp) and (397.77bp,186.55bp)  .. (eta4);
  \draw [<->] (f3) ..controls (361.25bp,144.52bp) and (399.74bp,161.64bp)  .. (431.0bp,149.0bp) .. controls (437.28bp,146.46bp) and (442.98bp,141.91bp)  .. (f4);
  \draw [<->] (f2) ..controls (223.75bp,136.72bp) and (245.14bp,145.34bp)  .. (265.0bp,149.0bp) .. controls (337.56bp,162.37bp) and (362.6bp,176.65bp)  .. (431.0bp,149.0bp) .. controls (437.28bp,146.46bp) and (442.98bp,141.91bp)  .. (f4);
\end{tikzpicture}
  \caption{
    Directed graph for a spatial item factor model with twelve response items
    ($\rv{Y}_j$), twelve auxiliary variables ($\rv{Z}_j$), four latent factors
    ($\rv{\theta}_k$), four Gaussian processes ($\gp_k$), four linear predictors
    ($\eta_k$), and six covariates ($x_l$).
  }
  \label{spifa_fig:dag_spifa}
\end{figure}

The relationship between covariates $\ve{x}(s)$, latent factors $\rve{\theta}(s)$,
auxiliary latent variables $\rve{Z}(s)$ and response variables $\rve{y}(s)$ can be seen
more clearly through an example of confirmatory SPIFA, as shown in \fig \ref{spifa_fig:dag_spifa}.
In this example some of the coefficients $\ve{a}_j$ are set to zero so that each factor is
only explained by a subset of items.
It can be seen that the 12-dimensional response vector $\rve{Y}(s)$ is reduced to a 4-dimensional space of factors $\rve{\theta}(s)$. These factors are allowed to be correlated with each other and also spatial correlation is permitted within factors. The top row in the figure shows how covariates $\ve{x}(s)$ are used to predict the latent factors $\rve{\theta}(s)$.

\subsection{Identifiability and restrictions}
\label{spifa_sub:identifiability_and_restrictions}

The model presented above is subject to the same \emph{identifiability} problems as those
found in factor analysis and structural equation modelling. Identifiability issues arise
when different sets of parameters lead to the same likelihood in a structured way - this
leads to symmetry in the posterior distribution in a Bayesian framework, (or objective
function in a classical approach), i.e. there are multiple modes. In our model, these
identifiability issues could be due to \emph{additive}, \emph{scaling}, \emph{rotational}
or \emph{reflection} aliasing, which will be discussed in detail below.  Although
sophisticated approaches can be taken to solve problems of identifiability like using the
priors proposed in \citet{Bhattacharya2011}, we use classical approaches as shown in
\citet{Geweke1996} given that inference of our model is computationally expensive
\citep[\se S2]{chacon-montalvan2025supplement}.

\textit{Additive} aliasing occurs when the item easiness parameter $c_j$ and the product
$\ve{a}_j^\intercal\ve{\theta}_i$ both have free means. In this situation, adding a
constant to one term and subtracting it from the other leaves the probability density
function unchanged. Similarly, if the item discrimination vector $\ve{a}^\intercal_j$ is
multiplied by a constant and the latent factor $\ve{\theta}_i$ is divided by the same
constant, the probability density function also remains unchanged—this is known as
\textit{scaling} aliasing. To address scaling and additive aliasing in exploratory IFA, as
defined in Equation \eqref{spifa_eq:ifa_aux_var}, it is common to assume that the latent
factors follow a standard normal distribution, $\ve{\theta}_{ik} \sim \df{N}(0, 1)$
\citep{Bafumi2005}. More generally, we can assume $\ve{\theta}_i \sim \df{N}(\ve{0},
\m{\Sigma}_\theta)$, and set $\m{\Sigma}_\theta = \m{I}$ for exploratory analysis, or
$\diag{\m{\Sigma}_\theta} = \ve{1}$ for confirmatory analysis.

The SPIFA model presented in \se \ref{spifa_sub:extension_to_the_spatial_domain} does not
suffer from additive aliasing, because the processes $\mgp(s)$ and $\ve{v}(s)$ are already
assumed to be zero-mean. As mentioned above, we also assume that the covariates included
in \eq \eqref{spifa_eq:def_abilities} are standardised, which implies that the latent
factor $\ve{\theta}(s)$ has mean zero. However, our model does suffer from scaling
aliasing, so it is necessary to restrict the variances of the latent factors
$\rv{\theta}_k(s)$. This is complicated by the presence of covariates and spatial
processes, because we cannot directly ensure that $\var{\ve{b}^\tr_k \ve{x}(s) + \gp_k(s)
+ v_k(s)} = 1$. One simple way to achieve the required restriction is to fix the variance
of one of the components in the latent factor structure given in \eq
\eqref{spifa_eq:def_abilities} \citep[see][\se S1]{chacon-montalvan2025supplement}.
This restriction is usually applied to the multivariate error term, $\ve{v}(s)$, as in
\citet{Tekwe2014}. It is sufficient to fix a diagonal matrix, $\m{D}$, containing the
marginal standard deviations of $\ve{v}(s)$, that is, $\diag{\m{D}} = (\sigma_{v_1},
\dots, \sigma_{v_m})^\tr$, such that $\m{\Sigma}_v = \m{D}\m{R}_v\m{D}$, where $\m{R}_v$
is a correlation matrix.

Although these restrictions change the interpretation of the discrimination parameters
$\ve{a}_j$, because the latent factors are no longer on a unit scale, they are necessary
only to make inference feasible. Therefore, a scaling transformation can be applied
post-estimation to recover the usual interpretation of the discrimination parameters:
\begin{align}
  \label{spifa_eq:scale_transform}
  \rv{Z}_j(s) = c_j + \ve{a}^\tr\m{Q}\m{Q}^{-1}\m{\theta}(s) +
  \epsilon_j(s),
\end{align}
where $\m{Q}$ is a diagonal matrix containing the standard deviations of $\ve{\theta}(s)$.
This transformation yields a new vector of latent abilities $\m{Q}^{-1} \ve{\theta}(s)$
with unit variances and rescaled discrimination parameters $\ve{a}^\tr \m{Q}$, restoring
the standard interpretation \citep[see][\se S2.4]{chacon-montalvan2025supplement}.

The other two types of aliasing, \textit{rotational} and \textit{reflection}, arise
because linear transformations of the discrimination parameters, ${\ve{a}_j^*}^\tr =
\ve{a}_j^\tr\m{\Lambda}^{-1}$, and of the latent factors, $\ve{\theta}_i^* =
\m{\Lambda}\ve{\theta}_i$, yield the same probability density function as the original
parameters $\ve{a}_j$ and $\ve{\theta}_j$ \citep{erosheva2011}. To resolve rotational
aliasing, $m(m-1)/2$ restrictions are needed to eliminate this non-uniqueness. The usual
approach is to set the discrimination matrix $\m{A}$ to be lower triangular in EIFA, or to
fix at least $m(m-1)/2$ elements to zero in CIFA \citep{Geweke1996}. For reflection
aliasing, there are $2^m$ equivalent matrices $\m{\Lambda}$, which arise by simultaneously
changing the signs of both $\ve{a}_j$ and $\ve{\theta}_i$. Identifiability can therefore
be ensured by fixing the signs of the diagonal elements of $\m{A}$ \citep{Geweke1996}.

For our SPIFA model, we require similar restrictions as described above. Focusing on the
confirmatory SPIFA setting, it is sufficient to fix at least $m(m-1)/2$ entries of $\m{A}$
to zero and to constrain one element in each column of $\m{A}$ to be strictly positive (or
negative). These restrictions can be imposed through a linear relationship between the
constrained parameters $\ve{a}_j^*$ and the free parameters $\ve{a}_j$:
\begin{equation}
  \label{spifa_eq:a_restr}
  \ve{a}^*_j = \ve{u}_j + \m{L}_j\ve{a}_j,
\end{equation}
where the vector $\ve{u}_j$ contains the fixed values, and the matrix $\m{L}_j$ indicates
which elements of the free parameter $\ve{a}_j$ remain active \citep{Cai2010a}. In the
example below, the third and fourth elements of the parameter vector are fixed to $0$ and
$1$, respectively.
\begin{equation}
  \left(
  \begin{array}[1]{c}
    a^*_{j1} \\
    a^*_{j2} \\
    a^*_{j3} \\
    a^*_{j4} \\
  \end{array}
  \right)\\
  =
  \left(
  \begin{array}[1]{c}
    0 \\
    0 \\
    0 \\
    1 \\
  \end{array}
  \right)\\ +
  \left(
  \begin{array}[4]{cccc}
    1 & 0 & 0 & 0\\
    0 & 1 & 0 & 0\\
    0 & 0 & 0 & 0\\
    0 & 0 & 0 & 0\\
  \end{array}
  \right)\\
  \left(
  \begin{array}[1]{c}
    a_{j1} \\
    a_{j2} \\
    a_{j3} \\
    a_{j4} \\
  \end{array}
  \right)\\
  =
  \left(
  \begin{array}[1]{c}
    a_{j1} \\
    a_{j2} \\
    0 \\
    1 \\
  \end{array}
  \right)\\
\end{equation}
In practice, under a Bayesian framework, the required positivity (or negativity)
constraints can be enforced by specifying appropriate marginal prior distributions
\citep[see][\se S2.2]{chacon-montalvan2025supplement}.

In theory, the SPIFA model can be used for both exploratory and confirmatory factor
analysis. However, we recommend applying it in a confirmatory setting, where no rotation
of the latent factors is performed. In this way, the correlation parameters remain
directly interpretable. By contrast, if the latent factors are rotated, as in exploratory
analysis, interpreting the correlation parameters becomes more challenging.

\subsection{Allowing further flexibility on the multivariate spatial structure}
\label{spifa_sub:allowing_further_flexibility_on_the_multivariate_spatial_structure}

In the discussion above, we proposed using a set of independent Gaussian processes in the
structure of the latent factors $\ve{\theta}(s)$.
It is adequate when the latent factors $\rv{\theta}_k(s)$ are spatially structured and independent.
However, in more general situations, it can be the case that some of the factors $\theta_k(s)$ are not spatially correlated, or that some of the unexplained variation in two or more factors may have a common (spatially-correlated) component. In this situation it will be desirable, respectively, to include spatial structure on only a subset of the factors, or to share the spatial structure across several factors.

In a similar way to how restrictions were imposed on the discrimination parameters, we can use an $m\times g$ transformation matrix $\m{T}$ to convert $g$ independent standard Gaussian processes in $\mgp(s)$ into an $m$-dimensional multivariate Gaussian process, $\mgp^*(s)$:
\begin{equation}
  \label{spifa_eq:mgp_restr}
  \mgp^*(s) = \m{T}\mgp(s).
\end{equation}
An example is given in \eq \eqref{spifa_eq:mgp_restr_example}, where after transforming, $\gp_1(s)$ is common to the first and second factor and the second factor has an additional spatial structure, namely $\gp_2(s)$; $\gp_3(s)$ features in the third factor, and the last factor does not include any Gaussian process i.e. it is not spatially structured.
\begin{equation}
  \label{spifa_eq:mgp_restr_example}
  \left(
  \begin{array}[1]{c}
    \gp^*_1(s) \\
    \gp^*_2(s) \\
    \gp^*_3(s) \\
    \gp^*_4(s) \\
  \end{array}
  \right)\\
  =
  \left(
  \begin{array}[3]{ccc}
    t_{11} & 0 & 0 \\
    t_{21} & t_{22} & 0 \\
    0 & 0 & t_{33} \\
    0 & 0 & 0 \\
  \end{array}
  \right)\\
  \left(
  \begin{array}[1]{c}
    \gp_1(s) \\
    \gp_2(s) \\
    \gp_3(s) \\
  \end{array}
  \right)\\
  =
  \left(
  \begin{array}[1]{c}
    t_{11}\gp_1(s) \\
    t_{21}\gp_1(s) + t_{22}\gp_2(s) \\
    t_{33}\gp_3(s) \\
    0 \\
  \end{array}
  \right)\\
\end{equation}
Notice that the variance of $\mgp^*(s)$ is controlled by $\m{T}$. Using this stochastic process in \eq \ref{spifa_eq:def_abilities}, we re-define the $m$-dimensional latent factor of our model as:
\begin{align}
  \label{spifa_eq:def_abilities_flex}
  \ve{\theta}(s) & = \m{B}^\tr\ve{x}(s)+ \mgp^*(s) + \ve{v}(s).
\end{align}

The methods described in the section are closely connected to multivariate geostatistical models of coregionalization \citep{Gelfand2004, Fanshawe2012}. The main difference is that, due to the sparse structure of $\m{T}$, it is not ensured that $\mgp^*(s)$ has a positive definite covariance matrix, and that this structure is a way for the user to control
the nature of interrelationships between factors (which would obviously change according to the problem and data under study), rather than allowing free reign estimating all the elements of $\m{T}$.  Although we propose $\mgp^*(s)$ because it is simple and easy to interpret, our model is not limited to this choice, it could be replaced with a more attractive multivariate stochastic process \citep[e.g. see][]{Gneiting2010}.

There is a sense in which the restrictions imposed can be thought of as prior specification. Provided the `correct' overall sparse structure of $\m{T}$ has been chosen, such restrictions are also beneficial; in particular if $m>g$ then inference becomes more tractable -- both in terms of computation, and subsequently interpretation.

\subsection{Auxiliary variables in the identifiable spatial item factor analysis}%
\label{spifa_sub:auxiliary_variables_identifiable_spifa}

Using the restricted discrimination parameters $\ve{a}^*$ defined in \eq
\eqref{spifa_eq:a_restr} and the new definition of the latent factor $\rve{\theta}(s)$ in
\eq \eqref{spifa_eq:def_abilities_flex}, we obtain an identifiable and flexible model for spatial item factor analysis where the auxiliary variables $\rv{Z}_j(s)$ have the following structure
\begin{align}
  \label{spifa_eq:def_auxiliary_marginal_restr}
  \rv{Z}_{j}(s) & = c_j + \ve{a}_j^{*\tr}\rve{\theta}(s) + \epsilon_{j}(s) =
  c_j + \ve{a}_j^{*\tr}[\m{B}^\tr\ve{x}(s) +
  \mgp^*(s) +
  \ve{v}(s)] + \epsilon_{j}(s).
\end{align}
We are assuming that the structure of the restricted discrimination parameters $\ve{a}^*_j$ and also the multivariate Gaussian process $\mgp^*(s)$ will be informed by expert opinion through direct involvement of researchers in the area of application and/or through consulting the academic literature in that area.

Doing this not only allows our model to be identifiable, but it also allows us to obtain interpretable latent factors which are practically useful to researchers in the field under consideration.

\subsection{Matrix form of the auxiliary variables}
\label{spifa_sub:matrix_form_of_the_auxiliary_veatiables}

Expressing the terms in our model at the individual level as above, and in \eq
\eqref{spifa_eq:def_auxiliary_marginal_restr}, is convenient for understanding the various
components. However, in the following, we will use the matrix form of our model in order
to define the likelihood function in \se \ref{spifa_sub:likelihood_function} and later
derive the conditional posterior distributions in the \sm{}
\citep[\se S2]{chacon-montalvan2025supplement}.

Before proceeding with the matrix form of our model, we introduce some further notational conventions. Let $\rve{\alpha}(s)$ be a $q$-variate random variable at spatial location $s$. Then if $\ve{s} = (s_1, s_2, \dots, s_n)^\tr$ is a set of locations, we will define the $q$-vector $\rve{\alpha}_i = \rve{\alpha}(s_i) = (\rv{\alpha}_1(s_i), \dots, \rv{\alpha}_q(s_i))^\tr$ and the $n$-vector $\rve{\alpha}_{[j]} = \rve{\alpha}_{j}(\ve{s}) = (\alpha_j(s_1), \dots, \alpha_j(s_n))^\tr$.

With the above conventions, the collection of auxiliary random variables $\rve{Z} = (\rve{Z}_{[1]}^\tr, \dots, \rve{Z}^\tr_{[q]})^\tr$ for $q$ items at $n$ locations can be expressed as
\begin{align}
  \label{spifa_eq:def_zvec_abilities}
  \rve{Z} & = (\m{I}_q\otimes\ve{1}_n)\ve{c} +
  (\m{A}^*\otimes\m{I}_n)\rve{\theta} +
  \ve{\epsilon}
\end{align}
where $\m{I}_q$ and $\m{I}_n$ are identity matrices of dimension $q$ and $n$ respectively, $\ve{1}_n$ is a $n$-dimensional vector with all elements equal to one, $\ve{c} = (c_1, \dots, c_q)^\tr$ is a vector arrangement of the easiness parameters, $\m{A}^*_{q\times m}=(\ve{a}^*_1, \dots, \ve{a}^*_q)^\tr$ is a matrix arrangement of the restricted discrimination parameters, $\ve{\theta} = (\ve{\theta}_{[1]}^\tr, \dots, \ve{\theta}^\tr_{[m]})^\tr$ and $\ve{\epsilon} = (\ve{\epsilon}^\tr_{[1]}, \dots, \ve{\epsilon}^\tr_{[q]})^\tr$ is a $nq$-vector of residual terms.

The vector of latent abilities $\rve{\theta}$ with respect to \eq \eqref{spifa_eq:def_abilities_flex} can be expressed as
\begin{align}
  \label{spifa_eq:def_abilities_vector}
  \rve{\theta} & = (\m{I}_m\otimes\ve{X})\ve{\beta} +
  (\m{T}\otimes\m{I}_n)\mgp +
  \ve{v},
\end{align}
where $\ve{\beta}=\mve{\m{B}}$ is a column-vectorization of the multivariate fixed
effects, $\m{X}_{n\times p}=(\ve{x}_1, \dots, \ve{x}_n)^\tr$ is the design matrix of the
covariates, $\mgp = (\mgp_{[1]}^\tr, \dots, \mgp^\tr_{[m]})^\tr$ is the collection of the
multivariate Gaussian process and $\ve{v}= (\ve{v}_{[1]}^\tr, \dots,
\ve{v}^\tr_{[m]})^\tr$ is the collection of the multivariate residual terms. Substituting
\eq \eqref{spifa_eq:def_abilities_vector} into \eq \eqref{spifa_eq:def_zvec_abilities}, we obtain:
\begin{align}
  \label{spifa_eq:def_zvec_marginal_restr}
  \rve{Z} & = (\m{I}_q\otimes\ve{1}_n)\ve{c} +
  (\m{A}^*\otimes\m{X})\ve{\beta} +
  (\m{A}^*\m{T}\otimes\m{I}_n)\mgp +
  (\m{A}^*\otimes\m{I}_n)\ve{v} +
  \ve{\epsilon}.
\end{align}
This matrix representation is useful for deriving the multivariate marginal and conditional distributions of $\rve{Z}$ in the following sections.

Alternatively, the collection of auxiliary variables $\rve{Z}$ can also be expressed as
\begin{align}
  \label{spifa_eq:def_zvec_discrimination}
  \rve{Z} & = (\m{I}_q\otimes\ve{1}_n)\ve{c} +
  (\m{I}_q\otimes\m{\Theta})\ve{a}^* +
  \ve{\epsilon} \nonumber \\
  & = (\m{I}_q\otimes\ve{1}_n)\ve{c} +
  (\m{I}_q\otimes\m{\Theta})\ve{u} +
  (\m{I}_q\otimes\m{\Theta})\m{L}\ve{a} +
  \ve{\epsilon},
\end{align}
where $\m{\Theta}_{n\times m} = (\rve{\theta}_{[1]}, \dots, \rve{\theta}_{[m]})$ is the
matrix of latent abilities, $\ve{u} = (\ve{u}_1^\tr, \dots, \ve{u}_q^\tr)^\tr$ are the
restrictions defined in \eq \eqref{spifa_eq:a_restr}, $\ve{a} = (\ve{a}_1^\tr, \dots,
\ve{a}_q^\tr)^\tr$ are the free discrimination parameters and $\m{L} =
\oplus_{j=1}^q\m{L}_j$ is the direct sum of the activation matrices defined in \eq
\eqref{spifa_eq:a_restr} (recall these link the free discrimination parameters $\ve{a}$
with the constrained discrimination parameters $\ve{a}^*$). We later use \eq \eqref{spifa_eq:def_zvec_discrimination} in the derivation of the conditional posterior distribution of the discrimination parameters $\ve{a}$.

\subsection{Likelihood function}
\label{spifa_sub:likelihood_function}

A challenging aspect of some motivating applications is the fact that not all items are observed for all subjects. More generally, it is common to have to deal with missing data (in this case item responses) in statistics, therefore in the present section we begin to introduce notation for observed and missing data; this will be revisited in
\citet[\se S2]{chacon-montalvan2025supplement}
and is also connected with prediction.

Let $\ve{s} = (s_1, s_2, \dots, s_n)^\tr$ be a set of locations at which data from $q$ items has been collected. Let the random variable $\rv{Y}_{ij} = \rv{Y}_j(s_i)$ be the $j$-th item response at location $s_i$. Using notation introduced in \se \ref{spifa_sub:matrix_form_of_the_auxiliary_veatiables}, let $\rve{Y} = (\rve{Y}_{[1]}^\tr, \dots, \rve{Y}_{[q]}^\tr)^\tr$ be the collection of responses to all items. These can be divided into two groups; the set of observed variables $\rve{Y}_{obs}$ and the set of variables that were missing $\rve{Y}_{mis}$.

The marginal likelihood function for our spatial item factor analysis model is obtained by integrating the joint density of the observed variables $\rve{Y}_{obs}$, the associated auxiliary variables $\rve{Z}_{obs}$ and the collection of latent abilities $\rve{\theta} = (\rve{\theta}_{[1]}, \dots, \rve{\theta}_{[m]})^\tr$;
\begin{align}
  \label{spifa_eq:likelihood_general1}
  \df{L}(\ve{c}, \ve{a}, \m{B}, \m{T}, \ve{\phi}, \m{R}_v)
  & =
  \int \int
  \pr{\ve{y}_{obs} \mid \ve{z}_{obs}}
  \pr{\ve{z}_{obs}, \ve{\theta} \mid \ve{c}, \ve{a}, \m{B}, \m{T},
  \ve{\phi}, \m{R}_v}
  \,d\ve{z}_{obs} \,d\ve{\theta},
\end{align}
where $\ve{a} = (\ve{a}_{1}^\tr, \dots, \ve{a}_{q}^\tr)^\tr$ is vector arrangement of all the discrimination parameters and $\ve{\phi} = (\phi_1, \dots, \phi_g)^\tr$ is the vector of scale parameters of the $g$-dimensional Gaussian process $\mgp(s)$. Note that the main computational cost inside the integral, $\mathcal{O}(n^3m^3)$, comes from evaluating the distribution associated with $\ve{\theta}$ which has a $mn \times mn$ covariance matrix.

In \eq \eqref{spifa_eq:likelihood_general1}, the structure of our model implies
\begin{align}
  \label{spifa_eq:likelihood_yobs}
  \pr{\ve{y}_{obs} \mid \ve{z}_{obs}} =
  \prod_{o_{ij} = 1} \pr{y_{ij} \mid z_{ij}},
\end{align}
where $o_{ij}$ is an indicator variable with value equals to one when the variable $\rv{Y}_{ij}$ has been observed (i.e. is not missing) and zero otherwise. We further have:
\begin{align}
  \label{spifa_eq:join_probaility_detail}
  \pr{\ve{z}_{obs}, \ve{\theta} \mid \ve{c}, \ve{a}, \m{B}, \m{T},
  \ve{\phi}, \m{R}_v}
  & =
  \prod_{o_{ij}=1}\pr{z_{ij}\mid \ve{\theta}_i, c_j, \ve{a}_j, \m{B}}
  \pr{\ve{\theta} \mid \m{T}, \ve{\phi}, \m{R}_v},
\end{align}
and variables on the right hand side are normally distributed.

Note that the definition of the likelihood function through \eq
\eqref{spifa_eq:likelihood_general1}, \eqref{spifa_eq:likelihood_yobs} and \eqref{spifa_eq:join_probaility_detail} does not depend on the missing observations. Therefore, if some items were not observed in some of the locations, inference will still be possible provided the missing data are \emph{missing at random} \citep{Merkle2011}.
Using this likelihood, inference from the model can proceed in a number of ways. Maximum
likelihood estimation can be achieved by approximating the likelihood function in \eq
\eqref{spifa_eq:likelihood_general1} using a variety of Monte Carlo methods or via
stochastic approximation \citep{Cai2010a}. However in the present article, we focus on a
Bayesian approach because of the simplicity and reliability of uncertainty computation
\citep{fox2010bayesian}.

Our likelihood function can also be written using the auxiliary variables associated with both the observed and missing responses:
\begin{align}
  \label{spifa_eq:likelihood_general2}
  \df{L}(\ve{c}, \ve{a}, \m{B}, \m{T}, \ve{\phi}, \m{R}_v)
  & =
  \int
  \pr{\ve{y}_{obs} \mid \ve{z}_{obs}}
  \pr{\ve{z} \mid \ve{c}, \ve{a}, \m{B}, \m{T},
  \ve{\phi}, \m{R}_v}
  \,d\ve{z}.
\end{align}
The advantage of this representation is that the joint density of the auxiliary variables
$\pr{\ve{z} \mid \ve{c}, \ve{a}, \m{B}, \ve{T}, \ve{\phi}, \m{R}_v}$  can be obtained in a
straightforward manner using \eq \eqref{spifa_eq:def_zvec_marginal_restr}. It is normally distributed with mean
\begin{equation}
  \label{spifa_eq:z_mean}
  \ve{\mu}_z = (\m{I}_q\otimes\ve{1}_n)\ve{c} +
  (\m{A}^*\otimes\m{X})\ve{\beta}
\end{equation}
and covariance matrix
\begin{equation}
  \label{spifa_eq:z_cov}
  \m{\Sigma}_z =
  (\m{A}^*\m{T}\otimes\m{I}_n)\m{\Sigma}_{\gp}(\m{T}^\tr\m{A}^{*\tr}\otimes\m{I}_n) +
  (\m{A}^*\otimes\m{I}_n)\m{D}\m{R}_{v}\m{D}(\m{A}^{*\tr}\otimes\m{I}_n)
\end{equation}
where $\m{\Sigma}_\gp = \oplus_{k=1}^g\m{\Sigma}_{\gp_k}$ is the direct sum of the
covariance matrices of the independent Gaussian processes. We prefer this last definition
of the likelihood function, as it facilitates handling missing data via data augmentation
\citep[\se S2.3]{chacon-montalvan2025supplement}.

\subsection{R package}
\label{spifa_sec:r_package}

Our SPIFA model is implemented in \texttt{C++} and can be used through our open-source
\texttt{R} package, \texttt{spifa}, available at
\url{https://github.com/ErickChacon/spifa}. The package implements the Bayesian inference
method described in \citet[\se S2]{chacon-montalvan2025supplement}, allowing users to
specify the structure of the multivariate Gaussian processes and prior hyperparameters;
model selection is also available via the DIC. The package includes functions for
summarising model output, performing MCMC diagnostics, and producing predictive maps using
\texttt{sf} methods \citep{pebesma2018}.


\section{Results}
\label{spifa_sec:predicting_food_insecurity_on_the_brazilian_amazonia}

In this section, we present results from our motivating application: the modelling and
prediction of food insecurity in a remote urban centre, Ipixuna, located in Amazonas
state, Brazil. The complete analysis, along with the underlying \texttt{R} code, is
available at \url{https://erickchacon.gitlab.io/food-insecurity-mapping}. Following our
analyses, we shared and discussed the results with local authorities and community
members to interpret the findings collaboratively and gain additional qualitative
insights. We also conducted site visits to neighbourhoods identified as particularly
vulnerable to observe their characteristics. These interactions were highly valuable for
interpreting the results.

\subsection{Data description}%
\label{spifa_sub:data_description}

The food insecurity questionnaire was conducted in August 2015 (low-water dry season) and
March 2016 (high-water rainy season) with 100 randomly sampled households per season.
Households were randomly selected, with sampling density adjusted based on the
number of households per census sector \citep{ibge2010instituto}.
Sampling points were geolocated using Open Street Map and Google Earth, and were restricted to the habitable area of Ipixuna, defined as a 20-meter radius from streets or river edges. We
approached the nearest household to each sample location for interviews and recorded the GPS coordinates of all households. This research received ethical approval from Brazil’s national health research ethics committee (CONEP-CNS; protocolo 45383215.5.0000.0005) and Lancaster University (S2014/126).
Our
questionnaire contains 18 items divided into 4 categories (Table \ref{tab:items}). Section
A refers to questions addressed to the households as a whole, Section B refers to adults
only, Section C to households with children, and Section D contains items related to the
regional context of Ipixuna. The full questionnaire is available in the \sm{}
\citep[\se S3]{chacon-montalvan2025supplement}.

The items with higher endorsement probability were numbers 15, 3,  1, 18, and 2, see Table
\ref{tab:items}. In the present context, endorsement simply means `answering with an
affirmative'. This indicates that it is common that Ipixuna citizens have to buy food on
credit (i.e. putting it on a 'tab' at a local shop), eat few food types, are worried that
food will end before they have means to acquire more, reduce meat or fish consumption, or
run out of food. Of the 200 surveyed households, 25 did not have children and this led to
missing data on the 6 items of associated with food insecurity in children, see Section C
in \tab \ref{tab:items}. This is treated as missing because it is desirable to obtain a
joint model for all the population.

\begin{table}[ht]
\centering
\caption{Summary of the food insecurity items: i) the number of missing values ($\#$NA) and the proportion of endorsement ($\pi$) are shown for the descriptive analysis, ii) the posterior median of the discrimination parameters $\{\m{\hat{A}}_{\cdot 1}, \m{\hat{A}}_{\cdot 2}, \m{\hat{A}}_{\cdot 3}\}$ are shown for the confirmatory factor analysis (CIFA), and iii) the posterior median of the discrimination and easiness parameters  $\{\m{\hat{A}}_{\cdot 1}, \m{\hat{A}}_{\cdot 2}, \m{\hat{A}}_{\cdot 3}, \hat{\ve{c}}\}$ are shown for the spatial item factor analysis (SPIFA).} 
\label{tab:items}
\begingroup\scriptsize
\begin{tabular}{cclccccccccr}
   \toprule[.15em] \multirow{2}[3]{*}{\textbf{Item}} & \multirow{2}[3]{*}{\textbf{Section}} & \multirow{2}[3]{*}{\textbf{Question}} & \multicolumn{2}{c}{\textbf{Descriptive}} & \multicolumn{3}{c}{\textbf{CIFA}} & \multicolumn{4}{c}{\textbf{SPIFA}} \\
 \cmidrule(rl){4-5} \cmidrule(rl){6-8} \cmidrule(rl){9-12}  & & & $\#$NA & $\pi$ & ${\m{\hat{A}}_{\cdot 1}}$ & ${\m{\hat{A}}_{\cdot 2}}$ & ${\m{\hat{A}}_{\cdot 3}}$ & ${\m{\hat{A}}_{\cdot 1}}$ & ${\m{\hat{A}}_{\cdot 2}}$ & ${\m{\hat{A}}_{\cdot 3}}$ & \multicolumn{1}{c}{$\hat{\ve{c}}$}  \\
 \midrule[.1em]1 & \multirow{3}[3]{*}{A} & worried that food ends &   0 & 0.56 & $\cdot$ & 1.63 & $\cdot$ & $\cdot$ & 1.77 & $\cdot$ & 0.44 \\ 
  2 &  & run out of food &   0 & 0.52 & $\cdot$ & $\cdot$ & 1.51 & $\cdot$ & $\cdot$ & 1.88 & 0.22 \\ 
  3 &  & ate few food types &   0 & 0.64 & 1.68 & $\cdot$ & $\cdot$ & 1.81 & $\cdot$ & $\cdot$ & 1.46 \\ 
   \hline
4 & \multirow{4}[3]{*}{B} & skipped a meal &   0 & 0.30 & 1.47 & $\cdot$ & 1.02 & 1.92 & $\cdot$ & 1.00 & -0.63 \\ 
  5 &  & ate less than required &   0 & 0.41 & 0.88 & $\cdot$ & 1.75 & 1.39 & $\cdot$ & 1.50 & -0.07 \\ 
  6 &  & hungry but did not eat &   0 & 0.24 & 1.26 & $\cdot$ & 1.49 & 1.81 & $\cdot$ & 1.54 & -1.44 \\ 
  7 &  & at most one meal per day &   0 & 0.26 & 1.57 & $\cdot$ & $\cdot$ & 1.80 & $\cdot$ & $\cdot$ & -0.55 \\ 
   \hline
8 & \multirow{6}[3]{*}{C} & ate few food types &  25 & 0.49 & 1.71 & $\cdot$ & $\cdot$ & 1.88 & $\cdot$ & $\cdot$ & 0.59 \\ 
  9 &  & ate less than required &  25 & 0.31 & 1.90 & $\cdot$ & $\cdot$ & 2.23 & $\cdot$ & $\cdot$ & -0.36 \\ 
  10 &  & decreased food quantity &  25 & 0.36 & 2.17 & $\cdot$ & $\cdot$ & 2.53 & $\cdot$ & $\cdot$ & -0.03 \\ 
  11 &  & skipped a meal &  25 & 0.23 & 1.99 & $\cdot$ & $\cdot$ & 2.52 & $\cdot$ & $\cdot$ & -1.05 \\ 
  12 &  & hungry but did not eat &  25 & 0.20 & 2.11 & $\cdot$ & $\cdot$ & 2.59 & $\cdot$ & $\cdot$ & -1.33 \\ 
  13 &  & at most one meal per day &  25 & 0.18 & 1.93 & $\cdot$ & $\cdot$ & 2.43 & $\cdot$ & $\cdot$ & -1.51 \\ 
   \hline
14 & \multirow{5}[3]{*}{D} & food just with toasted manioc flour &   0 & 0.17 & 0.34 & $\cdot$ & 1.25 & 0.63 & $\cdot$ & 1.28 & -1.62 \\ 
  15 &  & buying food on credit &   0 & 0.68 & $\cdot$ & 0.72 & $\cdot$ & $\cdot$ & 0.80 & $\cdot$ & 0.63 \\ 
  16 &  & borrowed food &   0 & 0.14 & $\cdot$ & 1.44 & $\cdot$ & $\cdot$ & 1.61 & $\cdot$ & -1.90 \\ 
  17 &  & had meals at neighbors &   0 & 0.17 & $\cdot$ & 0.98 & $\cdot$ & $\cdot$ & 1.04 & $\cdot$ & -1.26 \\ 
  18 &  & reduced meat or fish &   0 & 0.54 & 1.29 & $\cdot$ & $\cdot$ & 1.43 & $\cdot$ & $\cdot$ & 0.75 \\ 
   \bottomrule[.15em]\end{tabular}
\endgroup
\end{table}

\subsection{Confirmatory item factor analysis}%
\label{spifa_sub:confirmatory_item_factor_analysis}

Before undertaking the CIFA, we first performed an
EIFA to determine the number of dimensions and identify
which items should be related to each factor. We compared models whose dimensions ranging
from one to six and selected a model with three dimensions. A likelihood ratio test showed
no significant improvement for models with more than three dimensions ($p$-value $0.594$).
Similarly, the Akaike Information Criterion (AIC) increased by 17 units, and the Bayesian
Information Criterion (BIC) increased by 66 units when considering higher dimensions.
Based on these three criteria, models with two and three dimensions were most suitable,
and we choose to proceed with three dimensions. Finally, we applied a varimax rotation to
the selected model to inform the structure of the CIFA model.

For the structure of the CIFA model, we decided to include only those items with a
discrimination parameter greater than $0.5$ in the EIFA model. This results in the
following factor structure: the first factor is represented by items 3--14 and 18; the
second factor, by items 1 and 15--17; and the third factor, by items 2, 4--6, and 14. To
perform Bayesian inference, we used standard normal priors for the easiness parameters
$c_j$; standard normal priors for the discrimination parameters with exception of
$\{A_{11,1}, A_{13,1} A_{16,2}, A_{14,3}\}$ for which we used normal priors with mean $\mu
= 1$ and standard deviation $\sigma = 0.45$. To address reflection aliasing and ensure the
interpretability of our model, we employ a prior for these four parameters that constrains
them, with high probability, to be positive. This approach, explained in \se
\ref{spifa_sub:identifiability_and_restrictions}, ensures that the latent factors are
interpreted as measures of food insecurity rather than food security. We use a LKJ prior
distribution, as proposed in \citet[][\se 3]{Lewandowski2009}, with hyper-parameter $\eta
= 1.5$ for the correlation matrix of the latent factors. The adaptive MCMC scheme had
parameters $C = 0.7$ and $\alpha = 0.8$ with target acceptance probability of $0.234$.
This is a reference value which is an the optimal acceptance probability for the random
walk Metropolis-Hastings algorithm assuming a target normal distribution to balance
parameter space exploration and efficient convergence \citep{Andrieu2008}. We ran the
Metropolis-within-Gibbs algorithm described in
\citet[\se S2.3]{chacon-montalvan2025supplement} for $1.5$ million
iterations discarding the first 300,000 iterations and storing $1$ in $200$ of the
remaining iterations. Convergence was assessed using the improved split potential scale
reduction factor ($\hat{R}$) that compares the properties of the first half of a chain to
the second half \citep{vehtari2021ranknormalization}. All chains have $\hat{R}$ values
lower then 1.05 indicating convergence
\citep[see][\se S4.1]{chacon-montalvan2025supplement}.

The posterior median of the discrimination parameters $\{\m{\hat{A}}_{\cdot 1},
\m{\hat{A}}_{\cdot 2}, \m{\hat{A}}_{\cdot 3}\}$ of the CIFA model is shown in Table
\ref{tab:items}. These values show that questions related to the reduction of quality and
quantity of food in the diet of children (items 10--12) are the top three most important
items for the first factor, labeled as \textit{severe food insecurity including children}.
The second factor, labeled as \textit{anxiety and reliance on social relations}, includes
three items relating to Amazonian coping strategies (15--17) and one concerning anxiety
(1). Note that using credit (15), borrowing food (16), or relying on neighbors for meals
(17) are likely sources of anxiety in their own right. Finally, the third factor, labeled
as \textit{adults eating less and experiencing hunger}, is related mainly to the reduction
in quantity of food (2 and 4--6) and one item associated with the substitution of normal
foods with only toasted manioc (farinha) flour, a staple carbohydrate in low-income households (14).

\begin{figure}[!ht]
  \centering
  \includegraphics[width=\linewidth]{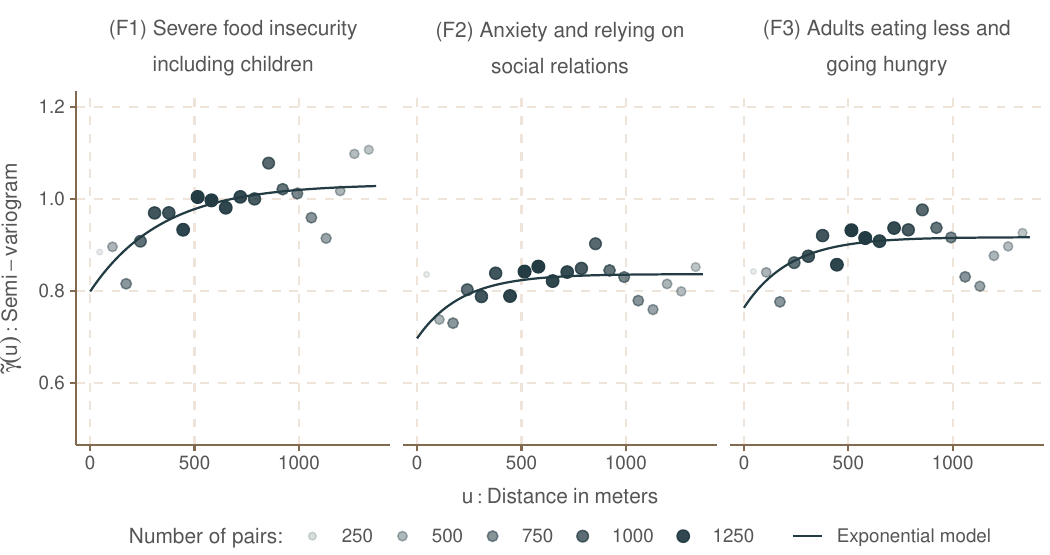}
  \caption{Empirical semi-variogram $\tilde{\gamma}(u)$ for each latent factor: points
  represent the empirical values, while the lines indicate fitted theoretical exponential
  models.}
  \label{spifa_fig:variogs}
\end{figure}

In order to evaluate the spatial correlation in the obtained factors, we use the empirical
semi-variogram: see Figure \ref{spifa_fig:variogs}. This exploratory tool for determining
the extent and form of (spatial) correlation is defined as a function of the distance $u$
between any two points,
\begin{equation*}
  \hat{\gamma}(u) = \frac{1}{2}\meanhat{(\gp(s) - \gp(s+u))^2}.
\end{equation*}
The initial increasing behaviour of the semi-variogram, mainly, observed in the first and
third factors is evidence for spatial correlation in these dimensions of food insecurity.

\subsection{Spatial confirmatory item factor analysis}%
\label{spifa_sub:spatial_confirmatory_item_factor_analysis}

We placed the same restrictions on the discrimination parameters for the spatial item factor as we did for the confirmatory item factor analysis. Based on the empirical variograms shown in \fig \ref{spifa_fig:variogs}, we proposed three models; model 1 includes a Gaussian process in the first latent factor (SPIFA I), model 2 includes Gaussian processes in the first and third factor (SPIFA II), and model 3 includes Gaussian processes in the three factors (SPIFA III).
In \fig \ref{spifa_fig:variogs}, we observe that the exponential model provides a good fit
to the empirical semi-variogram. For instance, for the first latent factor, it had the
lowest sum of squared errors ($\text{SSE} = 29.43$) compared to the Gaussian ($\text{SSE}
= 29.5$) and spherical ($\text{SSE} = 63.15$) models, and similar results were observed
for the other factors. Due to its superior fit and simplicity, we chose the exponential
correlation function, $\rho(u) = \exp(-u/\phi)$, to model the spatial structure of each
Gaussian process in our study.
Spatial predictors were not included in our model because these are insufficiently finely
resolved in our study area. For instance, there are only eight urban census sectors
in the 2010 demographic census
covering Ipixuna \citep{ibge2010instituto}. In the future, we are planning a larger scale analysis in which spatial predictors \emph{will} be included; our software package is already able to handle this case.

We used the same prior specifications as in the CIFA model for the easiness parameters $c_j$, discrimination parameters $A_{jk}$ and correlation matrix $\m{R}_v$. In addition, we used log-normal priors $\df{LN}(\log(160), 0.3)$, $\df{LN}(\log(80), 0.3)$ and $\df{LN}(\log(80), 0.3)$ for the scale parameters $\{\phi_1, \phi_2, \phi_3\}$ of the Gaussian processes in factor 1, 2 and 3 respectively; and the log-normal prior distribution
$\df{LN}(\log(0.4), 0.4)$ for all the free elements of $\m{T}$. The adaptive MCMC scheme
had parameters $C = 0.7$ and $\alpha = 0.8$ with target acceptance probability of $0.234$.
We ran the Metropolis-within-Gibbs algorithm described in
\citet[\se S2.3]{chacon-montalvan2025supplement}
for 1.5
million iterations discarding the
first 300,000 iterations and storing 1 in every 200 iterations.
Convergence was again assessed using the improved split potential scale reduction factor
($\hat{R}$). All chains of the three models have $\hat{R}$ values lower then 1.05
indicating convergence
\citep[see][\se S.4.1]{chacon-montalvan2025supplement}

We compared four models using the Deviance Information Criterion (DIC) proposed in
\citet{Spiegelhalter2002}, details of the computation for our model is provided in
the \sm{} \citep[\se S2.5]{chacon-montalvan2025supplement}.
In Table \ref{tab:dics}, we can see that the classical confirmatory model
(CIFA) has lowest
effective number of parameters ($329.82$); this model has independent random effects only.
In contrast, the spatial models include both independent and spatial random effects as
explained in \se \ref{spifa_sec:spatial_item_factor_analysis}. The DIC for the three
spatial models is lower than that for CIFA, hence by this measure, it is statistically
advantageous in terms of model fit to allow the factors to be spatially correlated. Among
the three spatial models, SPIFA III, which incorporates Gaussian processes in all three
factors, has the best performance in terms of DIC ($2198.8$). Consequently, the remainder
of this section focuses on the results from this model when applied to the \textit{full}
data set, the low-water \textit{dry} season data, and the high-water \textit{rainy} season
data.

\begin{table}[ht]
\centering
\caption{Deviance Information Criteria (DIC) for the Proposed Models: without spatial correlation (CIFA), with spatial correlation in factor 1 (SPIFA I), with spatial correlation in factor 1 and 3 (SPIFA II) and with spatial correlation in all factors (SPIFA III).} 
\label{tab:dics}
\begingroup\footnotesize
\begin{tabular}{lccc}
   \toprule[.1em] \multirow{2}[3]{*}{\textbf{Model}} & \multicolumn{3}{c}{\textbf{Diagnostics}} \\
 \cmidrule(rl){2-4} & Posterior Mean Deviance & Effective Number of Parameters & DIC \\
 \midrule[.05em] CIFA & 1894.31 & 329.82 & 2224.13 \\ 
  SPIFA I & 1866.67 & 335.14 & 2201.81 \\ 
  SPIFA II & 1861.22 & 338.64 & 2199.85 \\ 
  SPIFA III & 1857.54 & 341.26 & 2198.80 \\ 
   \bottomrule[.1em] \end{tabular}
\endgroup
\end{table}

\begin{figure}[!ht]
  \centering
  \includegraphics[width=1\linewidth, trim={0cm 0.3cm 0cm 0.3cm}]{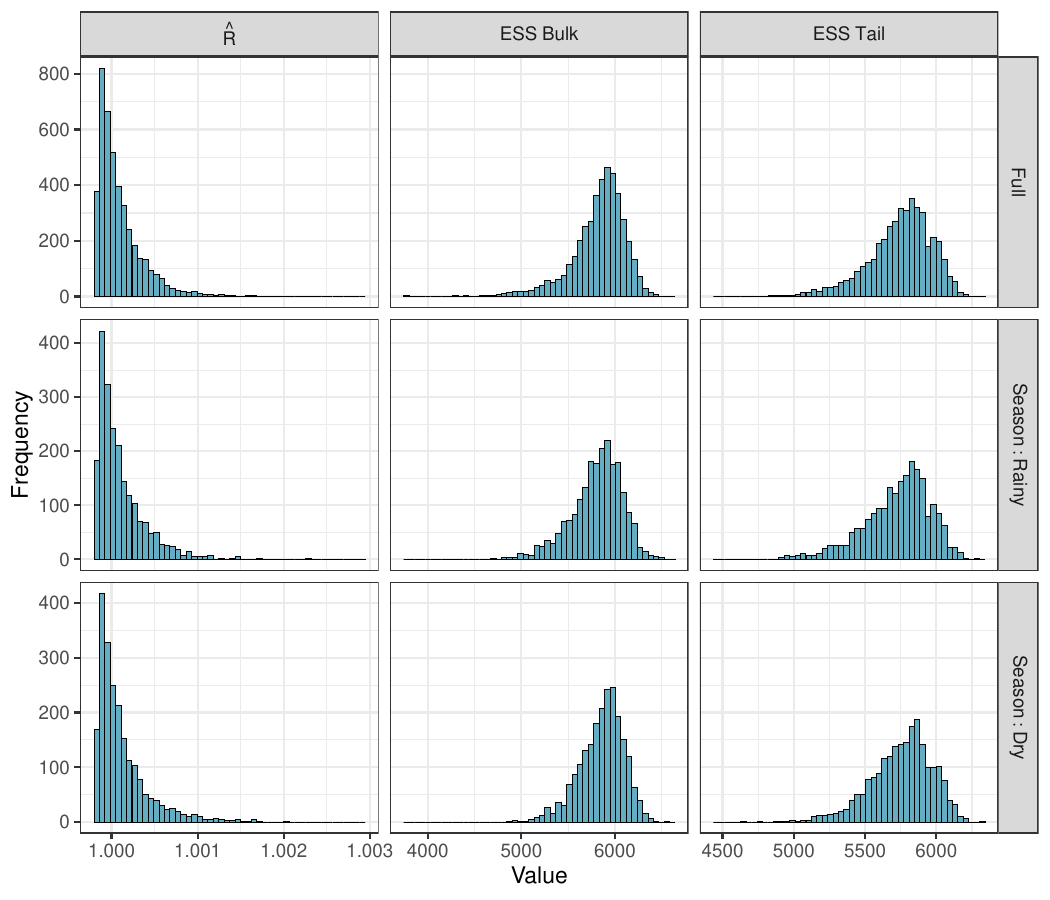}
  \caption{Histograms of the improved $\hat{R}$, bulk effective
  sample size (ESS Bulk), and tail effective sample size (ESS Tail) for assessing
  convergence across all MCMC chains for the SPIFA model applied to the full data, rainy
  season, and dry season.}
  \label{spifa_fig:spifa_diag}
\end{figure}

Diagnostics such as the improved $\hat{R}$, the bulk effective sample size (bulk-ESS), and
the tail effective sample size (tail-ESS) are presented in \fig \ref{spifa_fig:spifa_diag}
for the SPIFA model applied to the full, dry, and rainy season data sets. The bulk-ESS and
tail-ESS measure the effective sample size concerning the center and the tail of the
distribution, respectively \citep{vehtari2021ranknormalization}. The histograms of
$\hat{R}$ indicate that no chain had values greater than $1.05$, demonstrating convergence.
Traceplots for randomly selected chains can  be seen in the \sm{}
\citep[\se S4.2]{chacon-montalvan2025supplement}.
Additionally, the sample sizes, around $5900$, suggest a sufficiently large sample
size for reliable estimates of the mean, median, and credible intervals.

\subsubsection{Estimation of the discrimination and easiness parameters}

The posterior medians of the discrimination parameters $\{\m{\hat{A}}_{\cdot 1}, \m{\hat{A}}_{\cdot 2}, \m{\hat{A}}_{\cdot 3}\}$ for the selected model (SPIFA III) are shown in Table \ref{tab:items} under the column of SPIFA. We can see that the median of the obtained parameters have a broadly a similar structure as for the CIFA model, so their interpretation is as discussed in the previous section; notice that most of the discrimination parameters are higher for the SPIFA model. The last column of Table \ref{tab:items} shows the posterior median of the easiness parameters $\hat{\ve{c}}$; note the items with high easiness are those most frequently answered with an affirmative (`endorsed'). This column shows that eating few food types (item 3), obtaining credit for eating (item 15) and worrying that food will end (item 1) are the most common behaviours in the population of Ipixuna. Borrowing food (item 16), eating a meal containing only toasted manioc flour (item 14), only being able to feed your children one meal per day (item 13) and feeling hungry but do not eat (item 6 and 12) are less common.

\begin{figure}[!ht]
  \centering
  \includegraphics[width=1\textwidth]{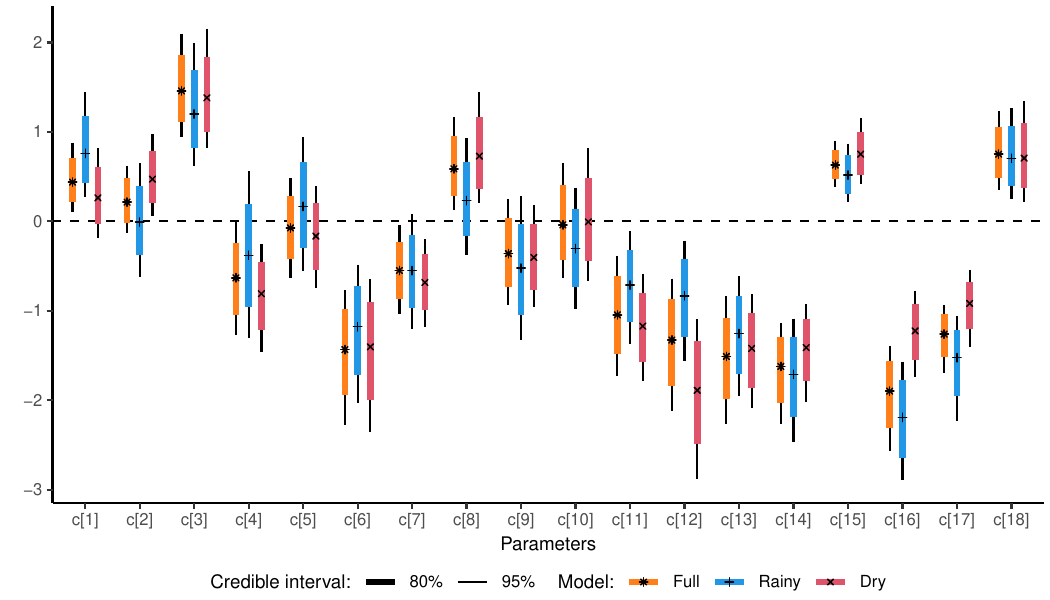}
  \caption{Credible intervals of the easiness parameters under three SPIFA models: using
  all data (Full), high-water season (Rainy) and low-water season (Dry). Intervals above
  (or below) zero for the easiness parameter indicate that the items are more (or less)
  likely then average to be endorsed.}
  \label{fig:ci_easiness}
\end{figure}

\begin{figure}[!ht]
  \centering
  \includegraphics[width=1\textwidth]{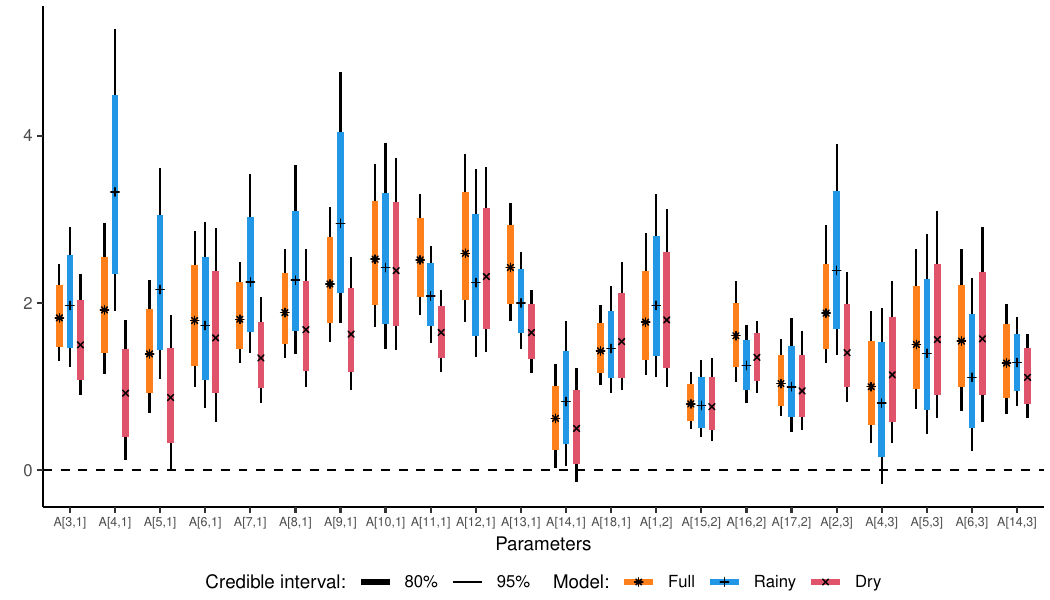}
  \caption{Credible intervals of the discrimination parameters $A[i,j]$ for item $i$ with
  respect to factor $j$ under three SPIFA models: all data (Full), high-water season
  (Rainy), and low-water season (Dry). Intervals above (or below) zero for the
  discrimination parameter indicates that the items are positively (or negatively) related
  with the factor of interest.}
  \label{fig:ci_discrimination}
\end{figure}

The comparison of the credible intervals for the easiness and discrimination parameters
for the full, rainy season, and dry season data is shown in Figures \ref{fig:ci_easiness}
and \ref{fig:ci_discrimination}, respectively. The estimates for the easiness parameters
are similar for most items, with some differences for item 12 (``hungry but did not eat'')
and item 16 (``borrowed food''). Item 12 is more commonly endorsed in the rainy season,
while item 16 is more commonly endorsed in the dry season. On the other hand, the
discrimination parameters also have similar values in both seasons, with the exception of
item 4 (``skipped a meal'') for the first latent factor, which has higher discrimination
in the rainy season. Although not significant, there seems to be a difference in the
discrimination of item 5 (``ate less than required'') and item 9 (``children ate less than
required'') with respect to the first factor, showing higher discrimination in the rainy
season.

\subsubsection{Spatial prediction of food insecurity}

\begin{figure}[!ht]
  \centering
  \includegraphics[width=1\linewidth, trim={0cm 0.3cm 0cm 0.3cm}]{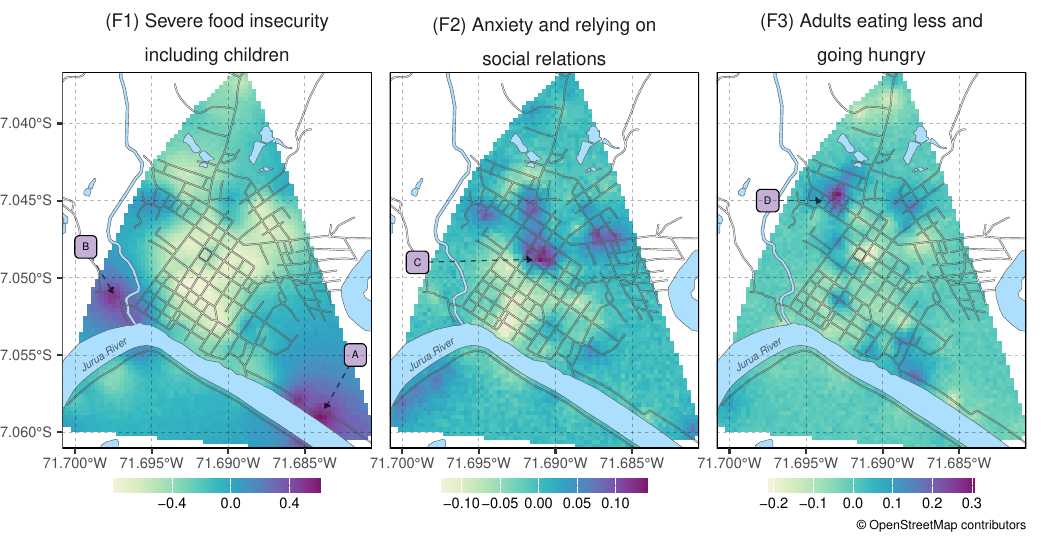}
  \caption{Median of the predicted spatial component of latent food insecurity using all
  data. Labeled regions indicate hotspots high levels of food insecurity.}
  \label{spifa_fig:sp_factors}
\end{figure}

The predicted median of the spatial component associated to each latent factor, when using
the full data set, is shown in \fig \ref{spifa_fig:sp_factors}. The left plot shows that
the first factor has a strong spatial structure; the respective posterior median of the
standard deviation and scale parameters of the associated Gaussian process are
$\hat{T}_{1,1} = 0.468~(95\%\text{CI}: 0.29 - 0.7)$ and $\hat{\phi}_1 = 216 ~\text{meters}
~(95\%\text{CI}: 127 - 365)$. Examining the middle plot, for the second factor, it can be
seen that the spatial structure is not as strong as the first factor. The respective
parameters of the associated Gaussian process have posterior medians $\hat{T}_{2,2} =
0.20~(\text{CI}: 0.09 - 0.41)$ and $\hat{\phi}_2 = 83~(\text{CI}: 45.1 - 148)$. The right
hand plot, referring to the third factor, shows moderate spatial structure with similar
median posterior estimates as the second factor: $\hat{T}_{3,3} = 0.29~(\text{CI}: 0.15 -
0.49)$ and $\hat{\phi}_3 = 76.8~(\text{CI}: 43.9 - 136)$.

\begin{figure}[!ht]
  \centering
  \includegraphics[width=1\linewidth, trim={0cm 0.3cm 0cm 0.3cm}]{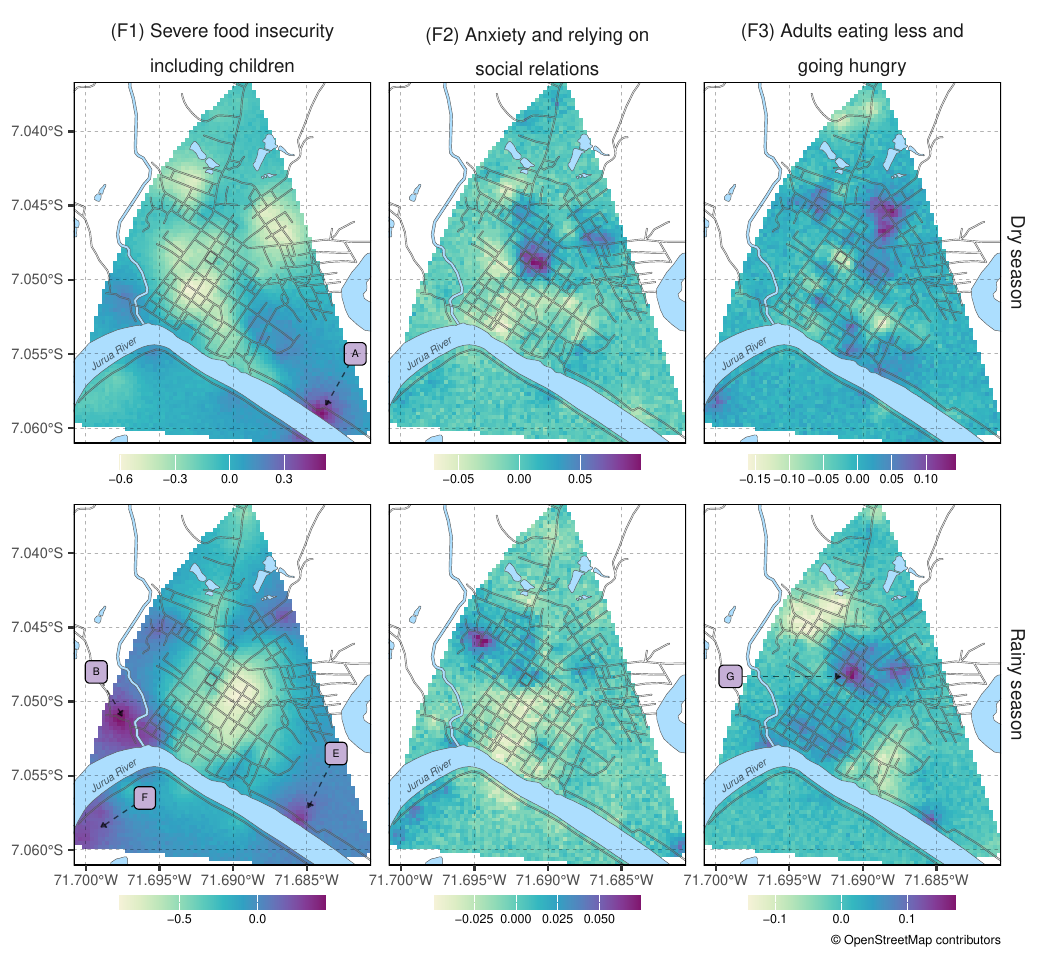}
  \caption{Median of the predicted spatial component of latent food insecurity for the
  high-water season (Rainy) and low-water season (Dry). Labeled regions are hotspots high
  levels of food insecurity.}
  \label{spifa_fig:sp_factors_season}
\end{figure}

\fig \ref{spifa_fig:sp_factors_season} shows the predicted median of the spatial component
associated with each latent factor when using the dry and rainy season data separately.
For the rainy season, the first component exhibits higher variability $(\hat{T}_{1,1} =
0.487)$ and a larger correlation scale parameter $(\hat{\phi}_1 = 197)$ compared to the
second $(\hat{T}{2,2} = 0.165, \hat{\phi}_2 = 80.7)$ and third factors $(\hat{T}_{3,3} =
0.295, \hat{\phi}_3 = 83.8)$. Similarly, for the dry season, the first component shows
greater variability $(\hat{T}_{1,1} = 0.435)$ and a higher correlation scale parameter
$(\hat{\phi}_1 = 178)$ than the second $(\hat{T}_{2,2} = 0.184, \hat{\phi}_2 = 79.6)$ and
third factors $(\hat{T}_{3,3} = 0.276, \hat{\phi}_3 = 76.2)$.

Spatial food insecure hotspots were identified as regions where the probability that the
spatial effect is positive is greater than $0.7$. Maps of these exceedance probabilities
are provided in
\citet[\se S4.4]{chacon-montalvan2025supplement}.
Using all data, we identified four hotspots (A, B, C,
D), as shown in \fig \ref{spifa_fig:sp_factors}. Under the same criteria, in the dry
season, only one hotspot (A) was identified, while in the rainy season, four hotspots (B,
E, F, G) were identified, as shown \fig \ref{spifa_fig:sp_factors_season}.

Examining the obtained maps of food insecurity for the first factor, we observe lower
levels of food insecurity around the center of the study area during both the rainy and
dry season, as well as when combining both seasons. When combining both seasons, more
severe food insecurity regions are found around locations A ($71.684\degrees$ W,
$7.059\degrees$ S) and B ($71.698\degrees$ W, $7.052\degrees$ S), as shown in \fig
\ref{spifa_fig:sp_factors}. Location A is also classified as hotspot in the dry season,
while location B is detected in the rainy season as well, as shown in \fig
\ref{spifa_fig:sp_factors_season}. Additionally, locations E ($71.686\degrees$ W,
$7.058\degrees$ S) and F ($71.698\degrees$ W, $7.059\degrees$ S) are only identified in
the rainy season.

In this city, location A is at the edge of the River Juru\'a.
It is a relatively new and very poor neighbourhood called Bairro da V\'arzea. Location B
refers to the flood-prone, politically marginalized and poor neighbourhood of Turruf\~ao.
Housing is on stilts, and there is no sanitation and poor provision of public services.
Location E is close to location A, also at the edge of the River Juru\'a, and is highly
flood-prone. Area F is another flood-prone peri-urban neighbourhood on the other side of
the River Juru\'a, by the name of Bairro da Ressaca. The common characteristic among these
locations is that they are poor, marginalized, and mostly flood-prone neighbourhoods on
the peri-urban fringe. Most of the heads of households in these neighbourhoods are
rural-urban migrants (often relatively recent), and many of their livelihoods are still
based in rural areas. The high levels of food insecurity in region A might not be due to
flood-risk but rather river-dependent livelihoods. All these relatively large areas, for
the first factor, indicate relatively severe food insecurity, yet without apparent anxiety
and a distinct absence of some coping strategies: borrowing food, eating in other
households or accessing credit.

With respect to the maps of the second factor, higher levels of food insecurity can be
identified around location C ($71.691\degrees$ W, $7.048\degrees$ S) when combining both
seasons (\fig \ref{spifa_fig:sp_factors}).
This location covers a large and older area of the town, including portions of two
neighbourhoods: Bairro do Cemet\'erio and Multir\~ao Velho. These areas are not
flood-prone and are not as marginalized or poor, though they are certainly not wealthy. It
is plausible that this factor captures more moderate food insecurity and coping strategies
associated with higher levels of horizontal social capital and access to credit.

In the maps of the third factor of food insecurity, we observe areas of severe food
insecurity around location D ($71.693\degrees$ W, $7.045\degrees$ S) when combining both
seasons, and around  G ($71.691\degrees$ W, $7.048\degrees$ S) in the rainy season.
Area D covers the border between two poor, peri-urban neighbourhoods: Bairro da Liberdade
and Multir\~ao Novo. Region G is similarly situated as location C, between the
neighborhoods Bairro do Cemet\'erio and Multir\~ao Velho, and shares similar
characteristics to C as explained above.

In general, the parameter estimates of our model do not vary drastically between the dry
and rainy seasons. However, we observe different patterns of food insecurity when
spatially predicting the latent factors. This indicates that it is relevant to acknowledge
that the levels of food insecurity in the population will vary depending on the season.
This is particularly important in cities like Ipixuna, where livelihoods are
river-dependent not only due to the risk of flooding but also because access to food
through fishing might be affected.

Identifying the areas of high (and also low) food insecurity is of relevance for future
research in this area, for example: exploring the social and environmental (e.g. household
flood risk due to elevation) determinants of vulnerability to food insecurity.
Understanding the spatial-variation of food insecurity at local (e.g. neighbourhood or
street) scales will also allow us to continue our dialogue with local government and other
stakeholders around which are the priority areas for intervention and what type(s) of
intervention should be deployed in order to reduce the risk of food insecurity.


\section{Discussion}
\label{spifa_sec:discussion}

\subsection{Spatial Item Factor Analysis}

In this work we have developed a new extension of item factor analysis to the spatial
domain, where the latent factors are allowed to be spatially correlated. Our model allows
for the inclusion of predictors to help explain the variability of the factors. These
developments allow us to make prediction of the latent factors at unobserved locations as
shown in our case of study of food insecurity in the Brazilian Amazon. We solved the
issues of identifiability and interpretability by employing a similar strategy as for
confirmatory item factor analysis in order to obtain an identifiable model, and by
standardizing the resulting factors after inference. Our model has been successfully
implemented in the \texttt{spifa} open source R package.
Since item factor analysis is used across such a wide range of scientific disciplines, we believe that our model and method of inference will be important for generating and investigating many new hypotheses. For instance, it could be used to model socio-economic status.

Computationally, our model is more efficient compared to a model where the spatial
structure is used at the level of the response variables. By including spatial structure
at the level of the factors, we reduce the computational cost from $\mathcal{O}(q^3n^3)$
to $\mathcal{O}(m^3n^3)$ where the number of items ($q$) is usually much greater than the
number of latent factors ($m$). For larger datasets, we can reduce the computational
burden by using alternatives to the Gaussian process. For example, we could use spatial
basis functions \citep{Fahrmeir2004}, nearest neighbour Gaussian processes
\citep{Datta2016} or stochastic partial differential equations \citep{Lindgren2011} to
reduce the cost. This is not so obvious because some of the nice properties of these
processes can be lost when working with multivariate models.

Our work can be extended in several directions. First, note that in our application we
adopted a two-stage modeling procedure: one stage to determine the sparse structure
between the latent factors and the response items, and a second stage to fit the SPIFA
model given that structure. In other applications, researchers may have prior knowledge to
define this structure directly, or the SPIFA model could be extended to stochastically
infer the sparse structure, thereby accounting for the associated uncertainty
\citep{rovckova2016fast}. In addition, our model can be extended to the spatio-temporal
domain, though again with increased computational expense, depending on the chosen
parameterisation of the spatio-temporal correlation. A more complex extension of our model
would allow the use of binary, ordinal and continuous items and would also allow
predictors to be related in a non-linear way to the latent factors. These extensions would
allow us to answer more complex research questions and would also improve prediction of
the latent factors \citep[see][\se S5]{chacon-montalvan2025supplement}.
Extensions to other distributional assumptions
(e.g. heavier tailed densities) are also possible if one desires to trade the convenience
conjugacy for realism; the Gaussian model fitted our particular dataset well.

\subsection{Food insecurity mapping in Ipixuna}

In our study of food insecurity in Ipixuna, Brazil, we demonstrated that the proposed
SPIFA models are more adequate than the classical CIFA models because food insecurity
contains a spatial component that may be explained by socio-economic conditions or common
environmental risks. In particular, the SPIFA model that introduces spatial correlation in
all the latent factors is more appropriate.

By fitting our model to the full data, rainy season data, and dry season data, we show
that the estimation of the parameters is consistent across both seasons. However, the
spatial variability differs between seasons. In the dry season, we identify a highly
food-insecure location (A) for the first factor at the edge of the River Juru\'a,
which is a relatively new and very poor neighbourhood called Bairro da V\'arzea. On the
other hand, in the rainy season, we identified three flood-prone regions with high levels
of food insecurity in the first factor: (B) the politically marginalized and poor
neighbourhood of Turruf\~ao, (E) a region close to hotspot A at the edge of the River
Juru\'a, and (F) a peri-urban neighbourhood on the other side of the River Juru\'a called
Bairro da Ressaca. Additionally, region G, situated between the neighborhoods Bairro do
Cemet\'erio and Multir\~ao Velho, was identified as highly food-insecure for the third
factor in the rainy season. It is plausible that this factor captures more moderate food
insecurity and coping strategies associated with higher levels of horizontal social
capital and access to credit.

Our results reinforce the idea that urban sub-populations in remote parts of Amazonas are
highly disadvantaged and experience food and nutrition insecurity likely due to the
following reasons:
(i) In Ipixuna and other small river-dependent towns in Amazonas, fishing enables poor urban households to eat more fish and may facilitate a modest reduction in their food insecurity risks. However, fishing is only partially effective in compensating for the socio-economic disadvantages experienced by these households \citep{rivero2022urban}.
(ii) The food security of urban households in Ipixuna is relatively dependent on what households can harvest or grow themselves in surrounding rural areas, or obtain through barter or market exchange with people in their social networks. For instance, someone in 45 percent of urban households in Ipixuna had been fishing in the previous 30 days
\citep{rivero2022urban}
and fish is the main animal-sourced food in such towns
\citep{carignanotorres2022wildmeat}.
Our data also show that 86 percent of urban households in Ipixuna had consumed bushmeat (terrestrial forest animals) in the previous year
\citep{torres2022rural}.
The accessibility of these agricultural products and wildfoods is highly seasonal and hence reliance on them - which is greater among the poorer peri-urban households -- is likely to be associated with certain food insecurity risks.
(iii) Overall, the nutritional status of children in Ipixuna and similar towns is precarious, with iron-deficiency anemia prevalence of 45 percent in children-under-five years old
\citep{carignanotorres2022wildmeat}.
(iv) The health inequities in Ipixuna and places like it are not the outcome of environmental variability. Instead, they reflect historical marginalization of Amazonia's riverine populations which has led to high vulnerability in these neglected remote cities
\citep{parry2018social}
and the invisibilization of climate-health risks facing marginalized populations
\citep{parry2019visible}.
The food security challenges of people in Ipixuna's poorest, peripheral neighbourhoods are, of course, influenced by measurable household metrics such as monetary income and family size. However, fundamentally, inequitable access to nutritious food across the urban population reflects power imbalances, politics, and policies. The precarious situation of household's in Turrufao is a consequence of multiple factors including: Statal neglect of rural populations (in-part because a lack of schools and healthcare drove a rural exodus and rapid urbanization); a failure of urban planning (in which the poorest people live in the most flood-prone areas, poorly served by basic amenities such as clean water and sanitation); the on-going neglect of the neighbourhood by urban elites
\citep[in some cases, the descendants of the former rubber bosses,][]{parry2019visible}
inter-mixed with corruption and governance failures; and deep multi-dimensional poverty, and social structures which perpetuate stigma and racism against indigenous people and non-tribal riberinhos of mixed ancestry.

\section*{Funding}

This research was funded by a Future Research Leaders Fellowship to LP (ES/K010018/1), the
Newton Fund/FAPEAM (ES/M011542/1), Brazil’s CNPq (CsF PVE 313742/2013-8) and
CAPES-ProAmazonia (Projeto 3322-2013), and the European Commission Horizon 2020 RISE
programme (Project 691053 - ODYSSEA).

\section*{Acknowledgments}

The authors would like to express their gratitude to the entire fieldwork team for their
efforts in collecting valuable data in the remote cities of the Brazilian Amazon,
particularly to Andre de Moraes for his outstanding work in data collection and his
contributions to the 2017 workshop discussions. We also extend our thanks to Prof. Naziano
Filizola for his invaluable collaboration during the workshops conducted with local
citizens and government officials.

\singlespacing
\bibliographystyle{apalike}
\bibliography{library}

\newpage
\renewcommand{\appendixpagename}{Supplementary Material}
\doublespacing
\begin{appendices}
  \renewcommand{\thesection}{S\arabic{section}}

\section{Spatial item factor analysis}
\label{spifa_app:spifa}

\subsection{Scaling aliasing}
\label{spifa_sub:scaling_aliasing}

Restricting the variances of the latent abilities to one, $\diag{\m{\Sigma}_{\theta}} = \ve{1}$, is the same as restricting the variances of the residual term $\ve{v}(s)$ because 
\begin{equation}
  \label{spifa_eq:variance_restriction}
  \var{\ve{b}^\tr_k \ve{x}(s) + \gp_k(s) + v_k(s)} = 1
\end{equation}
for $k = 1, \dots, m$ implies that
\begin{equation}
  \label{spifa_eq:residual_variance_restriction}
  \var{v_k(s)} = 1 - \var{\ve{b}^\tr_k \ve{x}(s) + \gp_k(s)} = \sigma^2_{v_k}.
\end{equation}
More generally, the constrain $\diag{\m{\Sigma}_{\theta}}$ is equivalent to set the covariance matrix of $\ve{v}(s)$ as $\m{\Sigma}_{v} = \m{D}_1\m{R}_v\m{D}_1$, where $\m{D}_1$ is a diagonal matrix with elements $\sigma_{v_k}$. Then the covariance matrix of the latent abilities $\rve{\theta}(s)$ is expressed as
\begin{equation}
  \label{spifa_eq:covariance_matrix_restricted}
  \var{\rve{\theta}(s)} =
  \var{\m{B}^\tr \ve{x}(s)} + \var{\mgp(s)} + \m{D}_1\m{R}_v\m{D}_1,
\end{equation}
the problem with this restriction is that $\sigma_{v_k}$ need to be known.

Inference can be attained by introducing arbitrary values. Consider the transformation $\m{D}_2=\m{D}\m{D}_1^{-1}$, where $\m{D}$ is a diagonal matrix with arbitrary values, then we can define
\begin{equation}
  \label{spifa_eq:scaling_aliasing_transform}
  \hat{\ve{a}}_j^\tr\rve{\hat{\theta}}(s) = \ve{a}_j^\tr\m{D}_2^{-1} \m{D}_2\rve{\theta}(s) = \ve{a}_j^\tr\rve{\theta}(s).
\end{equation}
Note that under this transformation, the variance of the new latent variable $\m{\hat{\theta}}(s) = \m{D}_2\rve{\theta}(s)$ is defined as
\begin{equation}
  \label{spifa_eq:covariance_matrix_restricted_hat}
  \var{\rve{\hat{\theta}}(s)} =
  \var{\hat{\m{B}}^\tr \ve{x}(s)} + \var{\hat{\mgp}(s)} + \m{D}\m{R}_v\m{D},
\end{equation}
where $\hat{\m{B}}^\tr = \m{D}_2\m{B}^\tr$ and $\hat{\mgp}(s) = \m{D}_2\mgp(s)$.
It can be seen that defining an arbitrary diagonal matrix $\m{D}$ still allows us to make inference given that the marginal variances of $\m{\hat{\theta}}(s)$ are still restricted. In this case, the variances are equal to the squared values of the diagonal matrix $\m{D}_2$, $\diag{\m{\Sigma}_{\hat{\theta}}} = \diag{\m{D}_2^2}$.

If we choose $\m{D} = \m{I}$; then $\m{D}_2 = \m{D}_1^{-1}$, $\m{\hat{\theta}}(s) = \m{D}_1^{-1}\rve{\theta}(s)$, $\diag{\m{\Sigma}_{\hat{\theta}}} = \diag{\m{D}_1^{-2}}$ and
\begin{equation}
  \label{spifa_eq:covariance_matrix_restricted_hat_iden}
  \var{\rve{\hat{\theta}}(s)} =
  \var{\hat{\m{B}}^\tr \ve{x}(s)} + \var{\hat{\mgp}(s)} + \m{R}_v.
\end{equation}
This transformation allows us to make inference, but the interpretation of the transformed parameters $\hat{\ve{a}_j}$ are not the same as in the classical item factor analysis because the marginal variances of $\hat{\rve{\theta}}(s)$ are not equal to $1$, $\diag{\Sigma_{\hat{\theta}}} \neq \ve{1}$. To recover the interpretation of the discrimination parameters, we simply compute the standard deviations of $\hat{\rve{\theta}}(s)$ after sampling to obtain the estimated $\m{Q} = \hat{\m{D}}_1^{-1}$, and back-transformed $\ve{a}_j=\m{Q}\hat{\ve{a}}_j$ and $\rve{\theta}(s) = \m{Q}^{-1}\rve{\hat{\theta}}(s)$ as explained in \se \ref{spifa_sub:scaling_samples_for_interpretation}.

%
%
%


\section{Bayesian Inference for the Spatial Item Factor Analaysis}
\label{spifa_sec:inference_using_markov_chain_monte_carlo}

In this section, we outline the Bayesian inference process for the spatial item factor
analysis model proposed in the main paper. We begin by presenting the Bayesian formulation
of our model in \se \ref{spifa_sub:bayes_spifa}. In \se \ref{spifa_sub:priors}, we detail
the prior specifications. The sampling scheme for the parameters and auxiliary variables,
using the Metropolis-within-Gibbs algorithm, is explained in \se
\ref{spifa_sub:sampling_scheme}. Post-sampling scaling procedures are described in \se
\ref{spifa_sub:scaling_samples_for_interpretation}, followed by a discussion on model
selection in \se \ref{spifa_sub:model_selection}. Finally, spatial prediction of the
latent factors is covered in \se \ref{spifa_sec:_prediction}.

\subsection{Bayesian Spatial Item Factor Analysis Model}
\label{spifa_sub:bayes_spifa}

We factorised the joint likelihood in an natural way into four levels. The first three
levels are: the data level $\pr{\ve{y}_{obs}\mid \ve{z}_{obs}}$, the auxiliary variable
level $\pr{\ve{z}\mid \ve{\theta}, \ve{c}, \ve{a}}$, and the latent factor level
$\pr{\ve{\theta} \mid \ve{\beta}, \m{T}, \ve{\phi}, \m{R}_v}$. For our Bayesian model, we
add an additional level for the prior distribution of the parameters $\ve{c}, \ve{a},
\ve{\beta}, \m{T}, \ve{\phi}$ and $\m{R}_v$. The posterior distribution of the model is
\begin{align}
  \label{spifa_eq:bayes_spifa_post}
  \pr{\ve{z}, \ve{c}, \ve{a}, \ve{\theta}, \ve{\beta}, \m{T}, \ve{\phi}, \m{R}_v
  \mid \ve{y}_{obs}}
  & \propto
  \pr{\ve{y}_{obs}\mid \ve{z}_{obs}}
  \pr{\ve{z}\mid \ve{\theta}, \ve{c}, \ve{a}}
  \pr{\ve{\theta} \mid \ve{\beta}, \m{T}, \ve{\phi}, \m{R}_v}
  \nonumber \\
  & \quad~
  \pr{\ve{c}}
  \pr{\ve{a}}
  \pr{\ve{\beta}}
  \pr{\m{T}}
  \pr{\ve{\phi}}
  \pr{\m{R}_v}.
\end{align}
This choice of factorisation allows us to take advantage of conjugacy for some parameters and also marginalise terms that may lead to slow convergence/mixing e.g. the multivariate Gaussian process $\mgp$ and the multivariate residual term $\ve{v}$.

\subsection{Priors}
\label{spifa_sub:priors}

We assume Gaussian distributions for the easiness, discrimination and fixed effects parameters:
\begin{align}
  \label{spifa_eq:prior_items_params}
  \ve{c} & \sim \df{N}(\ve{\mu}_c, \m{\Sigma}_c), &
  \ve{a}_j & \sim \df{N}(\ve{\mu}_{a_j}, \m{\Sigma}_{a_j}), &
  \ve{\beta} & \sim \df{N}(\ve{0}, \m{\Sigma}_{\beta}),
\end{align}
where $\ve{\mu}_c$ and $\ve{\mu}_{a_j}$ are the mean parameters, and $\m{\Sigma}_c$, $\m{\Sigma}_{a_j}$ and $\m{\Sigma}_{\beta}$ are diagonal covariance matrices.

With respect to the $m$-dimensional Gaussian process $\mgp^*(s)$, we assume that the associated parameters have a log-normal distribution,
\begin{align}
  \label{spifa_eq:prior_gp_params}
  \text{vec}^*(\m{T}) & \sim \df{LN}(\ve{\mu}_T, \m{\Sigma}_{T}), &
  \ve{\phi} & \sim \df{LN}(\ve{\mu}_{\phi}, \m{\Sigma}_{\phi}),
\end{align}
where $\text{vec}^*(\m{T})$ is a vector of the non-zero values of $\m{T}$, $\m{\mu}_T$ and $\m{\mu}_{\phi}$ are the mean parameters and $\m{\Sigma}_T$ and $\m{\Sigma}_{\phi}$ are diagonal covariance matrices of the log-transformation of the parameters.

Finally, as proposed in \citet[][\se 3]{Lewandowski2009} we use an LKJ distribution for the correlation matrix $\m{R}_v$ of the multivariate residual term, which is defined as:
\begin{align}
  \label{spifa_eq:prior_lkj_corr}
  \pr{\m{R}_v} \propto \det(\m{R}_v)^{\eta - 1}.
\end{align}
Here, $\eta$ is the shape parameter of the LKJ distribution. If $\eta=1$, the density is uniform; for bigger values $\eta>1$,  the mode is a identity matrix; and band diagonal matrices are more likely when $0<\eta<1$.

Bayesian inference can be sensitive to the choice of hyperparameters for small sample sizes on the prior distributions described above; however, this is less highlighted in factor models due to the high number of observations $nq$. In our experience, inference does not vary drastically for the prior distribution of the easiness, discrimination and fixed parameters as long as reasonable hyperparameters are defined. More careful specification is needed for the scale parameters of the Gaussian processes. This can be achieved by using the maximum spatial distance between observations to define more informative prior distributions for these parameters.

\subsection{Sampling Scheme}%
\label{spifa_sub:sampling_scheme}

Samples from the posterior distribution in \eq \eqref{spifa_eq:bayes_spifa_post} are drawn using blocked Gibbs sampling where possible. In cases where the conditional posterior distribution is not available analytically, we use Metropolis Hastings to update parameters, details below.

\subsubsection{Auxiliary variables}
\label{spifa_ssub:auxiliary_variables}

Recall from above that we introduced a distinction between the observed variables $\rve{Y}_{obs}$ and the set that could not been observed $\rve{Y}_{mis}$. In a similar way, we divide the associated auxiliary variables into two groups, $\rve{Z}_{obs}$ and $\rve{Z}_{mis}$.
The joint vector of auxiliary variables $\rve{Z}$ is normally distributed given the easiness parameters $\ve{c}$, the discrimination parameters $\ve{a}$ and the latent factors $\rve{\theta}$:
\begin{align}
  \label{spifa_eq:z_joint_df}
  \pr{\ve{z} \mid \ve{c}, \ve{a}, \ve{\theta}}
  & = \df{N}(\ve{z} \mid
  (\m{I}_q\otimes\ve{1}_n)\ve{c} + (\m{A}^*\otimes\m{I}_n)\ve{\theta}, \m{I}_{nq}).
\end{align}
In the equation above it can be seen that any two elements of $\rve{Z}$ are conditionally independent given $\ve{c}$, $\ve{a}$ and $\rve{\theta}$ because the covariance is the identity matrix. Using the fact that this joint density can also be written as the product of two marginal densities and that $\rve{Y}_{obs}$ is conditionally independent of $\rve{Z}_{mis}$ given $\rve{Z}_{obs}$. The conditional posterior distribution for the auxiliary variables $\pr{\ve{z} \mid \ve{y}_{obs}, \ve{c}, \ve{a}, \ve{\theta}}$ is
\begin{align}
  \label{spifa_eq:post_z_joint}
  \pr{\ve{z}_{obs}, \ve{z}_{mis} \mid \ve{y}_{obs}, \ve{c}, \ve{a}, \ve{\theta}}
  & \propto
  \pr{\ve{z}_{obs} \mid \ve{c}, \ve{a}, \ve{\theta}}
  \pr{\ve{z}_{mis} \mid \ve{c}, \ve{a}, \ve{\theta}}
  \pr{\ve{y}_{obs}\mid \ve{z}_{obs}}.
\end{align}
Hence
the conditional posterior distribution for $\rve{Z}_{obs}$ is
\begin{align}
  \label{spifa_eq:post_z_obs}
  \pr{\ve{z}_{obs} \mid \ve{y}_{obs}, \ve{c}, \ve{a}, \ve{\theta}}
  & \propto
  \pr{\ve{z}_{obs} \mid \ve{c}, \ve{a}, \ve{\theta}}
  \prod_{o_{ij} = 1} \pr{y_{ij} \mid z_{ij}},
\end{align}
which is a marginal truncated normal distribution obtained from \eq \eqref{spifa_eq:z_joint_df}. Note that $\pr{{y}_{ij}\mid {z}_{ij}} = \ind{z_{ij}>0}^{y_{ij}}\ind{z_{ij}\leq 0}^{1-y_{ij}}$, where $\ind{.}$ is the indicator function. In a similar way, we obtain that the conditional posterior distribution of the auxiliary variables related to the missing data $\rve{Z}_{mis}$ as
\begin{align}
  \label{spifa_eq:post_z_mis}
  \pr{\ve{z}_{mis} \mid \ve{y}_{obs}, \ve{c}, \ve{a}, \ve{\theta}}
  & \propto
  \pr{\ve{z}_{mis} \mid \ve{c}, \ve{a}, \ve{\theta}},
\end{align}
which is a marginal distribution of \eq \eqref{spifa_eq:z_joint_df}. Hence, the only difference between the posterior of  both sets of variables is that it is truncated for the $\rve{Z}_{obs}$ and unrestricted for $\rve{Z}_{mis}$.

\subsubsection{Latent Factors}
\label{spifa_ssub:latent_factors}

The conditional posterior distribution of the latent abilities is
\begin{equation}
  \pr{\ve{\theta} \mid \ve{z}, \ve{c}, \ve{a}, \ve{\beta}} \propto
  \pr{\ve{z} \mid \ve{c}, \ve{a}, \ve{\theta}}
  \pr{\ve{\theta} \mid \ve{\beta}, \m{T}, \ve{\phi}, \m{R}_v},
\end{equation}
where the joint density of the auxiliary variables $\pr{\ve{z} \mid \ve{c}, \ve{a}, \ve{\theta}}$ is a Gaussian distribution
, given in \eq \eqref{spifa_eq:z_joint_df}, and the density of the latent factors
is also a Gaussian distribution,
\begin{equation}
\pr{\ve{\theta} \mid \ve{\beta}, \m{T}, \ve{\phi}, \m{R}_v} =
  \df{N}(\ve{\theta}\mid (\m{i}_m\otimes\m{x})\m{\beta},
  (\m{t}\otimes\m{i}_n)\m{\Sigma}_{\gp}(\m{t}^\tr\otimes\m{i}_n) +
  \m{D}\m{R}_v\m{D}\otimes\m{I}_n),
\end{equation}
where $\m{\Sigma}_\gp = \oplus_{k=1}^g\m{\Sigma}_{\gp_k}$.
Hence, the conditional posterior $\pr{\ve{\theta} \mid \ve{z}, \ve{c}, \ve{a}, \ve{\beta}}$ is defined by the product of two normal densities that leads to a normal density with covariance matrix
\begin{align}
  \label{spifa_eq:postcon_theta_cov}
  \m{\Sigma}_{\theta\mid\cdot}
  & =
  \left(
    (\m{a}^{*\tr}\otimes\m{i}_n)
    (\m{a}^{*}\otimes\m{i}_n) +
    \left(
      (\m{t}\otimes\m{i}_n) \m{\Sigma}_{\gp} (\m{t}^\tr\otimes\m{i}_n) +
      \m{D}\m{R}_v\m{D}\otimes\m{I}_n
    \right)^{-1}
  \right)^{-1},
\end{align}
and mean
\begin{align}
  \label{spifa_eq:postcon_theta_mean}
  \ve{\mu}_{\theta\mid\cdot}
  & =
  \m{\Sigma}_{\theta\mid\cdot}(\m{a}^{*\tr}\otimes\m{i}_n)
  (\ve{z} - (\m{I}_q\otimes\ve{1}_n)\ve{c}) +\nonumber \\
  & \quad~
  \m{\Sigma}_{\theta\mid\cdot}
  \left(
    (\m{t}\otimes\m{i}_n) \m{\Sigma}_{\gp} (\m{t}^\tr\otimes\m{i}_n) +
    \m{D}\m{R}_v\m{D}\otimes\m{I}_n
  \right)^{-1}
  (\m{i}_m\otimes\m{x})\m{\beta}.
\end{align}

\subsubsection{Fixed effects}
\label{spifa_ssub:fixed_effects}

For the multivariate fixed effects $\ve{\beta}$, the conditional posterior
\begin{equation}
  \label{spifa_eq:postcon_beta}
  \pr{\ve{\beta} \mid \ve{y}_{obs}, \ve{z}, \ve{c}, \ve{a}} \propto
  \pr{\ve{\theta} \mid \ve{\beta}, \m{T}, \ve{\phi}, \m{R}_v}
  \pr{\ve{\beta}}
\end{equation}
is given by the product of two normal densities,
\begin{align}
  \df{N}(\ve{\theta}\mid (\m{i}_m\otimes\m{x})\m{\beta},
  (\m{t}\otimes\m{i}_n)\m{\Sigma}_{\gp}(\m{t}^\tr\otimes\m{i}_n)+\m{R}\otimes\m{I}_n)
  \df{N}(\ve{\beta}\mid \ve{0}, \m{\Sigma}_{\beta}),
\end{align}
that also leads to a multivariate normal distribution with covariance matrix
\begin{align}
  \label{spifa_eq:postcon_beta_cov}
  \m{\Sigma}_{\beta\mid\cdot}
  & =
  \left(
    (\m{i}_m\otimes\m{X})^\tr
    \left(
      (\m{t}\otimes\m{i}_n) \m{\Sigma}_{\gp} (\m{t}^\tr\otimes\m{i}_n) +
      \m{R}\otimes\m{I}_n
    \right)^{-1}
    (\m{i}_n\otimes\m{x}) +
    \m{\Sigma}_{\beta}^{-1}
  \right)^{-1},
\end{align}
and mean
\begin{align}
  \label{spifa_eq:postcon_beta_mean}
  \ve{\mu}_{\beta\mid\cdot}
  & =
  \m{\Sigma}_{\beta\mid\cdot}(\m{i}_m\otimes\m{X}^\tr)
  \left(
    (\m{t}\otimes\m{i}_n) \m{\Sigma}_{\gp} (\m{t}^\tr\otimes\m{i}_n) +
    \m{R}\otimes\m{I}_n
  \right)^{-1}
  \ve{\theta}.
\end{align}

\subsubsection{Easiness parameters}
\label{spifa_ssub:easiness_parameters}

The conditional posterior distribution of the easiness parameters $\ve{c}$,
\begin{equation}
  \label{spifa_eq:postcon_easiness}
  \pr{\ve{c} \mid \ve{y}, \ve{z}, \ve{a}, \ve{\theta}} \propto
  \pr{\ve{z}\mid \ve{\theta}, \ve{c}, \ve{a}}
  \pr{\ve{c}},
\end{equation}
is also the product of two normal densities obtained from \eq \eqref{spifa_eq:z_joint_df}
and \eqref{spifa_eq:prior_items_params},
\begin{align}
  \label{spifa_eq:postcon_c}
  \pr{\ve{c} \mid \ve{y}, \ve{z}, \ve{a}, \ve{\theta}}
  & \propto
  \df{N}(\ve{z} \mid
  (\m{I}_q\otimes\ve{1}_n)\ve{c} + (\m{A}^*\otimes\m{I}_n)\ve{\theta}, \m{I}_{nq})
  \df{N}(\ve{c}\mid \ve{0}, \m{\Sigma}_{c}),
\end{align}
leading to a multivariate normal density with covariance matrix
\begin{align}
  \label{spifa_eq:postcon_c_cov}
  \m{\Sigma}_{c\mid\cdot}
  & =
  ((\m{I}_q\otimes\ve{1}_n)^\intercal(\m{I}_q\otimes\ve{1}_n)
  + \m{\Sigma}_c^{-1})^{-1} = (\diag{\m{\Sigma}_c}^{-1} + n)^{-1},
\end{align}
and mean
\begin{align}
  \label{spifa_eq:postcon_c_mean}
  \ve{\mu}_{c\mid\cdot}
  & =
  \m{\Sigma}_{c\mid\cdot}
  (\m{I}_q\otimes\ve{1}_n^\intercal)(\ve{z}-(\m{A}^*\otimes\m{I}_n)\ve{\theta}).
\end{align}

\subsubsection{Discrimination parameters}
\label{spifa_ssub:discrimination_parameters}

Due to the structure of our hierarchical model in \se \ref{spifa_sub:bayes_spifa}, the conditional posterior distribution of the discrimination parameters,
\begin{equation}
  \label{spifa_eq:postcon_discrimination}
  \pr{\ve{a} \mid \ve{y}, \ve{z},\ve{c}, \ve{\theta}} \propto
  \pr{\ve{z}\mid \ve{\theta}, \ve{c}, \ve{a}}
  \pr{\ve{a}},
\end{equation}
is determined by the product of two Gaussian densities obtained from \eq
\eqref{spifa_eq:z_joint_df} and \eqref{spifa_eq:prior_items_params},
\begin{align}
  \label{spifa_eq:postcon_a}
  \df{N}(\ve{z} \mid
  (\m{I}_q\otimes\ve{1}_n)\ve{c} +
  (\m{I}_q\otimes\m{\Theta})\ve{u} +
  (\m{I}_q\otimes\m{\Theta})\m{L}\ve{a},
  \m{I}_{n})
  \df{N}(\ve{a}\mid \ve{\mu}_a, \m{\Sigma}_{a}),
\end{align}
which, similar to previous parameters, leads to a Gaussian density with covariance matrix
\begin{align}
  \label{spifa_eq:postcon_a_cov}
  \m{\Sigma}_{a\mid\cdot}
  &
  = (\m{L}^\tr(\m{I}_q\otimes{\bm{\Theta}}^\tr\bm{\Theta})\m{L}
  + \m{\Sigma}_{a}^{-1})^{-1},
\end{align}
and mean
\begin{align}
  \label{spifa_eq:postcon_a_mean}
  \ve{\mu}_{a\mid\cdot}
  &
  = \m{\Sigma}_{a\mid\cdot}\m{l}^\tr{(\m{i}_q\otimes\m{\Theta}_j^{\tr})}
  (\ve{z} -
  (\m{I}_q\otimes\ve{1}_n)\ve{c} -
  (\m{I}_q\otimes\m{\Theta})\ve{u}) +
  \m{\Sigma}_{a\mid\cdot} \m{\Sigma}_{a}^{-1} \ve{\mu}_a .
\end{align}

\subsubsection{Covariance parameters}
\label{spifa_ssub:covariance_parameters}

Unlike the previous parameters, the parameters of the multivariate Gaussian process $\mgp^*(s)$ and the multivariate residual term $\ve{v}(s)$ can not be directly sampled from their conditional posterior density as they are not available analytically. However, this density can be defined up to a constant of proportionality,
\begin{equation}
  \label{spifa_eq:postcon_cov_pars}
  \pr{\text{vec}^*(\m{T}), \ve{\phi}, \m{R}_v \mid \ve{\theta}, \ve{\beta}} \propto
  \pr{\ve{\theta} \mid \ve{\beta}, \m{T}, \ve{\phi}, \m{R}_v}
  \pr{\m{T}}
  \pr{\ve{\phi}}
  \pr{\m{R}_v}.
\end{equation}
In order to obtain an MCMC chain that mixes over the real line, we work with $\log(\ve{\phi})$ instead of $\rve{\phi}$ and $\log(\text{vec}^*(\m{T}))$ instead of  $\text{vec}^*(\m{T})$. For the correlation $\m{R}_v$, we use canonical partial correlation, transforming to a set of free parameters $ \ve{\nu} \in \real^{m(m-1)/2}$, see \citet{Lewandowski2009} for further details.

We use an adaptive random-walk Metropolis Hastings algorithm to sample from this part of the posterior distribution. The covariance matrix of the proposal, is adapted to reach a fixed acceptance probability (e.g. 0.234). More specifically, we implemented algorithm 4 proposed in \citet{Andrieu2008} using a deterministic adaptive sequence $\gamma_i = C / i ^ \alpha$ for $\alpha \in ((1+\lambda)^{-1}, 1]$, where $\lambda > 0$. In the tests we have run and in our food insecurity application, this algorithm and choice of parameters performs well (see details below for our choice of $C$ and $\alpha$).

\subsection{Scaling Samples for Interpretation}%
\label{spifa_sub:scaling_samples_for_interpretation}

In \se 2.4 of the main paper, we saw how restricting the standard deviations of the
multivariate residual term $\ve{v}(s)$ is necessary to make our model identifiable.
However, we can not ensure that the latent factors will be on the same scale, which leads to a loss of interpretation of the discrimination parameters $\ve{a}_j$. As proposed in the same section, after the samples of the MCMC have obtained, we can transform the parameters in order to obtain latent factors with expected variance equal to 1 to solve this problem. We can then obtain the matrix $\m{Q}$
by filling the diagonal elements with the expected variances of the samples of the latent factors $\rve{\theta}(s)$. We then make the following transformations
\begin{align}
  \label{spifa_eq:scaleing_abilities}
  \ve{a}_j & \leftarrow \m{Q}\ve{a}_j, &
  \rve{\theta}_i & \leftarrow \m{Q}^{-1}\m{\theta}_i, &
  \m{B} & \leftarrow \m{Q}^{-1}\m{B}, &
  \m{T} & \leftarrow \m{Q}^{-1}\m{T}, &
  \m{D} & \leftarrow \m{Q}^{-1}\m{D};
\end{align}
the correct interpretation of the parameters is then recovered.

\subsection{Model Selection Using the Deviance Information Criterion}%
\label{spifa_sub:model_selection}

Bayesian model selection for the spatial item factor analysis can be performed using
various in information criteria commonly applied in Bayesian modelling
\citep[see][]{gelman2014understanding}. Here we focus on the \textit{Deviance Information
Criterion} (DIC) proposed by \citet{Spiegelhalter2002}, though we acknowledge competing
alternatives like the Watanabe-Akaike Information Criterion (WAIC). DIC,
a Bayesian counterpart to
the \textit{Akaike Information Criterion} (AIC), encapsulates the trade-off between
model fit and complexity. This complexity, measured through the effective
number of parameters, is determined by the difference between the mean of the deviance and
the deviance of the mean,
\begin{equation}
  p_D = \overline{D(\ve{\alpha})} - D(\bar{\ve{\alpha}}).
\end{equation}
The deviance in our case is given by
\begin{equation}
  D(\ve{\alpha}) = -2 \log\{\pr{\ve{y}\mid \ve{\alpha}}\}
  + 2 \log \{\pr{\ve{y} \mid \mu(\ve{\alpha}) = \ve{y}}\},
\end{equation}
where $\pr{\ve{y} \mid \mu(\ve{\alpha}) = \ve{y}}$ is the likelihood associated with a saturated model. The $DIC$ can then be calculated as:
\begin{equation}
  DIC = \overline{D(\ve{\alpha})} + p_D,
\end{equation}
where models with a lower $DIC$ are preferred. As evident from these expressions, the DIC
relies on the prior distribution and the convergence of the MCMC chains. The DIC is known
to perform well when the posterior distribution is adequately summarized by its mean
\citep{gelman2014understanding}, which is the case for our spatial item factor analysis.

In order to be able to calculate this quantity for our model, we  require the density function of the responses $\rve{Y}$ given all the parameters of the model, which is expressed as
\begin{equation}
  \log(\pr{\ve{y}\mid \ve{\alpha}}) =
  \sum_{o_{ij}=1}
  \left(y_{ij} \log(\Phi(c_j + \ve{a}_j^\tr\ve{\theta}_j)) +
  (1-y_{ij})\log(1-\Phi(c_j + \ve{a}_j^\tr\ve{\theta}_j))\right),
\end{equation}
where $o_{ij}$ is a binary variable taking value equal to one when the variable $Y_{ij}$ has been observed and zero otherwise.

\subsection{Prediction of Latent Factors}%
\label{spifa_sec:_prediction}

In this section, our interest is on the spatial prediction of the latent factors $\tilde{\rve{\theta}}$ at a set of locations that we have not observed data, $\tilde{\ve{s}}$. As is customary, we obtain the predictive distribution by integrating out the parameters of the model from the joint density of $\tilde{\rve{\theta}}$ and the parameters,
\begin{align}
  &
  \pr{\tilde{\ve{\theta}} \mid \ve{y}, \m{X}, \tilde{\m{X}}} =
  \nonumber \\
  &
  \int
  \pr{\tilde{\ve{\theta}} \mid
  \ve{\theta}, \m{B}, \ve{\sigma}^2, \ve{\phi}, \m{R}_v, \ve{y}, \m{X}, \tilde{\m{X}}}
  \pr{\ve{\theta}, \m{B}, \ve{\sigma}^2, \ve{\phi}, \m{R}_v \mid \ve{y}, \m{X}}
  d\ve{\theta}
  d\ve{B}
  d\ve{\sigma}^2
  d\ve{\phi}
  d\ve{\m{R}}_v.
  \nonumber
\end{align}

Note that a vectorized version of
the latent factors
can be expressed as
\begin{align}
  \label{spifa_eq:def_abilities_vector_pred}
  \tilde{\ve{\theta}} = (\m{I}_m \otimes \tilde{\m{X}})\ve{\beta} +
  (\m{T}\otimes\m{I}_{\tilde{n}})\tilde{\mgp} + \tilde{\ve{v}}.
\end{align}
Under these expressions, it can be shown that $\rve{\theta}$ and $\tilde{\rve{\theta}}$ are normally distributed with parameters
\begin{align}
  \label{spifa_eq:def_abilities_vector_mean}
  \mean{\ve{\theta}} & = (\m{I}_m\otimes\m{X})\ve{\beta}, &
  \var{\ve{\theta}} & =
  (\m{T}\otimes\m{I}_{n})\var{\mgp}(\m{T}^\tr\otimes\m{I}_{n}) + \var{\ve{v}}\\
  \mean{\tilde{\ve{\theta}}} & = (\m{I}_m\otimes\tilde{\m{X}})\ve{\beta}, &
  \var{\tilde{\ve{\theta}}} & =
  (\m{T}\otimes\m{I}_{\tilde{n}})\var{\tilde{\mgp}}(\m{T}^\tr\otimes\m{I}_{\tilde{n}})
  + \var{\tilde{\ve{v}}}.
\end{align}
Furthermore, the cross-covariance can  be obtained as
\begin{align}
  \label{spifa_eq:def_abilities_vector_cov}
  \cov{\tilde{\ve{\theta}}}{\ve{\theta}} & =
  \cov{(\m{T}\otimes\m{I}_{\tilde{n}})\tilde{\mgp} +
  \tilde{\ve{v}}}{(\m{T}\otimes\m{I}_{n})\mgp + \ve{v}}
  \nonumber \\
  & =
  \cov{(\m{T}\otimes\m{I}_{\tilde{n}})\tilde{\mgp}}{(\m{T}\otimes\m{I}_{n})\mgp}
  \nonumber \\
  & =
  (\m{T}\otimes\m{I}_{\tilde{n}})\cov{\tilde{\mgp}}{\mgp}(\m{T}^\tr\otimes\m{I}_{n}),
\end{align}
where $\cov{\tilde{\mgp}}{\mgp}$ is a block diagonal matrix as both $\tilde{\mgp}$ and
$\mgp$ are multivariate independent Gaussian process, see \se 2.3 of the main paper.
Hence, the conditional distribution of the latent abilities $\tilde{\ve{\theta}}$ is
$\pr{\tilde{\ve{\theta}} \mid \ve{\theta}, \m{B}, \ve{\sigma}^2, \ve{\phi}, \m{R}_v, \ve{y}, \m{X}, \tilde{\m{X}}}$, a normal distribution with mean and variance
\begin{align}
  \label{spifa_eq:ddfff}
  \mean{\tilde{\ve{\theta}} \mid \ve{\theta}}
  & =
  \mean{\tilde{\ve{\theta}}} + \cov{\tilde{\theta}}{\theta}\var{\theta}^{-1}
  (\ve{\theta} - \mean{\ve{\theta}}))
  \\
  \var{\tilde{\ve{\theta}} \mid \ve{\theta}}
  & =
  \var{\tilde{\ve{\theta}}} -
  \cov{\tilde{\theta}}{\theta}\var{\theta}^{-1}\cov{\theta}{\tilde{\theta}}.
\end{align}
Predictions are obtained by generating $\tilde{\ve{\theta}}$ from this conditional distribution for a set of samples $\ve{\theta}, \m{B}, \ve{\sigma}^2, \ve{\phi}, \m{R}_v$ obtained from the joint posterior via MCMC.
Obtaining the spatial prediction of only the spatial process $\pr{\tilde{\ve{\mgp}} \mid
\ve{y}, \m{X}, \tilde{\m{X}}}$ is similarly done.


\section{Full Questions Used in Survey}
\label{spifa_app:questions}

In this section, we give an explanation and translation of the 18 questions used in our `Food Insecurity Survey of Road-less Urban Areas and Surrounding Rural Areas in Amazonas (2015-16)' by researchers from Lancaster University, the Oswaldo Cruz Foundation (FioCruz) and the Federal Universities of Par\'a (UFPA) and Amazonas (UFAM).

The following questions were used in interviews with heads of households regarding their perceptions of food insecurity. These questions are based on the Brazilian Food Insecurity Scale \citep{segall-correa2014}, though modified to reflect a 30 day (rather than 3-month) time period and with slight changes in wording to reflect the fact that cash is not the only means to acquire food in the Amazonian context. Note that the English is a `back-translation' of what was really asked.

\subsection*{Section A:}

\emph{Portuguese: Nos \'ultimos 30 dias, ou seja, desde o dia \rule{2cm}{0.4pt} (mesmo dia atual) do m\^es de \rule{2cm}{0.4pt} (1 m\^es atr\'as):}
English: During the past 30 days:
\vspace{1em}

\noindent\textbf{Question 1:}
\emph{Portuguese: Voc\^es, deste domic\'ilio, j\'a tiveram a preocupa\c{c}\~ao de que os alimentos acabassem antes de poderem comprar ou receber mais comida?}
English: Were you, in this household, worried that you would run out of food before being able to buy or receive more food?

\noindent\textbf{Question 2:}
\emph{Portuguese: Os alimentos acabaram antes que voc\^es tivessem condi\c{c}\~oes para adquirir mais comida?}
English: Did you run out of food before having the means to acquire more?

\noindent\textbf{Question 3:}
\emph{Portuguese: Voc\^es comeram apenas alguns poucos tipos de alimentos que ainda tinham, porque o dinheiro acabou?}
English: Did you have to consume just a few types of foods (remaining) because you ran out of money?

\subsection*{Section B:}

\emph{Portuguese: Agora vou perguntar apenas sobre voc\^e e os outros adultos (18 anos ou mais) da sua casa. Algum de voc\^es, alguma vez:}
English: Now I'm going to ask you only about you and other adults (18 years and above) in your household. Did any of you adults:
\vspace{1em}

\noindent\textbf{Question 4:}
\emph{Portuguese: Deixou de fazer alguma refei\c{c}\~ao, porque n\~ao havia dinheiro para comprar comida?}
English: Skip a meal because there was not enough money to buy food?

\noindent\textbf{Question 5:}
\emph{Portuguese: Comeu menos do que achou que devia, porque n\~ao havia dinheiro para comprar comida?}
English: Eat less than what you thought you should because there was not enough money to buy food?

\noindent\textbf{Question 6:}
\emph{Portuguese: Sentiu fome, mas n\~ao comeu porque n\~ao havia dinheiro para comprar comida?}
English: Feel hungry but did not eat because there was not enough money to buy food?

\noindent\textbf{Question 7:}
\emph{Portuguese: Fez apenas uma refei\c{c}\~ao ao dia ou ficou um dia inteiro sem comer, porque n\~ao havia dinheiro para comprar a comida?}
English: Go without eating for a whole day or just have one meal in a whole day because there was not enough money to buy food?

\subsection*{Section C:}

\emph{Portuguese: Agora vou perguntar apenas sobre os moradores menores de 18 anos da sua casa. Algum deles, alguma vez:}
English: Now I'm going to ask only about those in the household under 18 years old. Did any of them:
\vspace{1em}

\noindent\textbf{Question 8:}
\emph{Portuguese: Comeu apenas alguns poucos tipos de alimentos que ainda tinham, porque o dinheiro acabou?}
English: Eat only a few types of food that you still had left, because money had run out?

\noindent\textbf{Question 9:}
\emph{Portuguese: N\~ao comeu quantidade suficiente de comida porque n\~ao havia dinheiro para comprar comida?}
English: Not eat enough because there was not enough money to buy food?

\noindent\textbf{Question 10:}
\emph{Portuguese: Foi diminu\'ida a quantidade de alimentos das refei\c{c}\~oes de algum morador com menos de 18 anos de idade, porque n\~ao havia dinheiro para comprar a comida?}
English: Reduce the size of meals of your children/adolescents because there was not enough money to buy food?

\noindent\textbf{Question 11:}
\emph{Portuguese: Deixou de fazer alguma refei\c{c}\~ao, porque n\~ao havia dinheiro para comprar comida?}
English: Skip a meal because there was not enough money to buy food?

\noindent\textbf{Question 12:}
\emph{Portuguese: Sentiu fome, mas n\~ao comeu porque n\~ao havia dinheiro para comprar mais comida?}
English: Were your children/adolescents ever hungry but you just could not buy more food?

\noindent\textbf{Question 13:}
\emph{Portuguese: Fez apenas uma refei\c{c}\~ao ao dia ou ficou sem comer por um dia inteiro, porque n\~ao havia dinheiro para comprar comida?}
English: Did your children go without food for a whole day or just have one meal in a whole day because there was not enough money to buy food?

\subsection*{Section D -- Regionalized food security questions}

\emph{Portuguese: Nos \'ultimos 30 dias, ou seja, desde o dia \rule{2cm}{0.4pt} do m\^es passado, alguma vez, o(a) senhor(a) ou algu\'em aqui desta casa:}
English: During the previous 30 days, at some time did you or anyone else in this household:
\vspace{1em}

\noindent\textbf{Question 14:}
\emph{Portuguese: Fez alguma refei\c{c}\~ao apenas com farinha ou chib\'e porque n\~ao tinha outro alimento?}
English: Had a meal with only toasted manioc flour (or this with water and salt) because there were no other foods?

\noindent\textbf{Question 15:}
\emph{Portuguese: Teve que pegar cr\'edito ou comprar fiado na taberna, mercadinho ou vendedor para comprar comida porque n\~ao tinha mais dinheiro?}
English: Have to borrow money or buy food on credit at a shop because there was no other money?

\noindent\textbf{Question 16:}
\emph{Portuguese: Emprestou comida de outra fam\'ilia porque faltou em casa e n\~ao tinha dinheiro?}
English: Borrowed food from another Family because you had none at home and had no money?

\noindent\textbf{Question 17:}
\emph{Portuguese: Fez as refei\c{c}\~oes na casa de vizinhos, amigos ou parentes porque n\~ao tinha comida em casa?}
English: Had meal(s) in the home of neighbours, friends or relatives because there was no food at home?

\noindent\textbf{Question 18:}
\emph{Portuguese: Diminuiu a quantidade de carne ou peixe em alguma refei\c{c}\~ao para economizar?}
English: Reduce the quantity of meat or fish in a meal in order to economize?


\section{Predicting Food Insecurity in an Urban Centre in Brazilian Amazonia}
\label{spifa_app:fim}

\subsection{MCMC diagnostics for the CIFA and SPIFA models}

\begin{figure}[H]
  \centering
  \includegraphics[width=\linewidth]{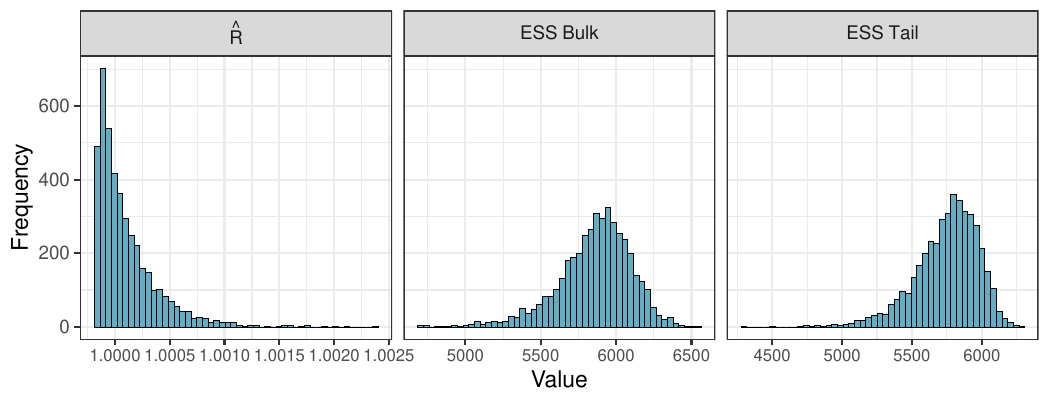}
  \caption{Histograms of the improved $\hat{R}$, bulk effective
  sample size (ESS Bulk), and tail effective sample size (ESS Tail) across all MCMC chains for the CIFA model.}
  \label{spifa_fig:cifa_diagnostics}
\end{figure}

\begin{figure}[H]
  \centering
  \includegraphics[width=\linewidth]{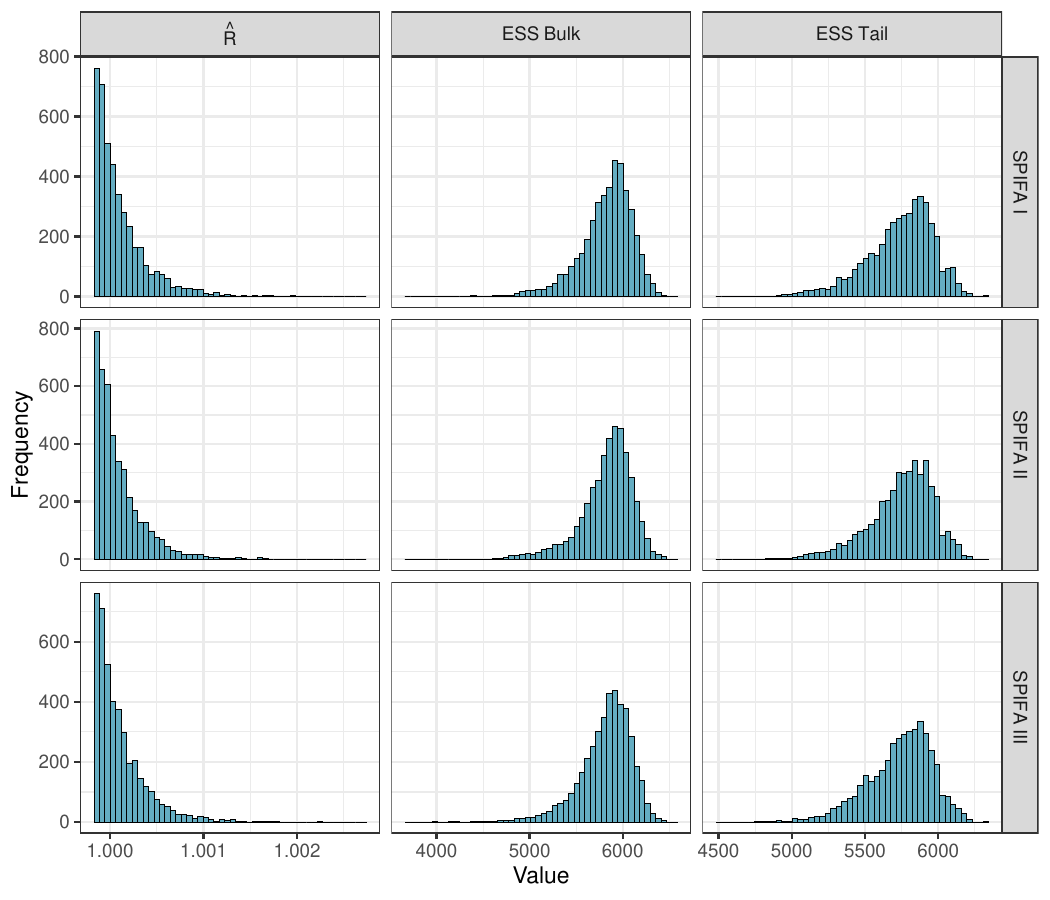}
  \caption{Histograms of the improved $\hat{R}$, bulk effective
  sample size (ESS Bulk), and tail effective sample size (ESS Tail) across all MCMC chains
  for the SPIFA models.}
  \label{spifa_fig:spifas_diagnostics}
\end{figure}

Figures \ref{spifa_fig:cifa_diagnostics} and \ref{spifa_fig:spifas_diagnostics} display histograms of the improved $\hat{R}$, bulk effective sample size (bulk-ESS), and tail effective sample size (tail-ESS) for the confirmatory item factor analysis (CIFA) and spatial item factor analysis (SPIFA) models, respectively. The results indicate that the $\hat{R}$ values for all chains are below 1.05, demonstrating convergence. The sample sizes, approximately 5900, suggest a sufficiently large dataset to provide reliable estimates of the mean, median, and credible intervals.

\subsection{MCMC traceplots for the SPIFA III model}
\label{spifa_app:traceplots}


In this section, we present trace plots of randomly selected parameters for each component
of our selected model (SPIFA III). These visualizations confirm the diagnostic results,
showing that the chains have achieved convergence across all parameters.

\begin{figure}[H]
  \centering
  \includegraphics[width=\linewidth]{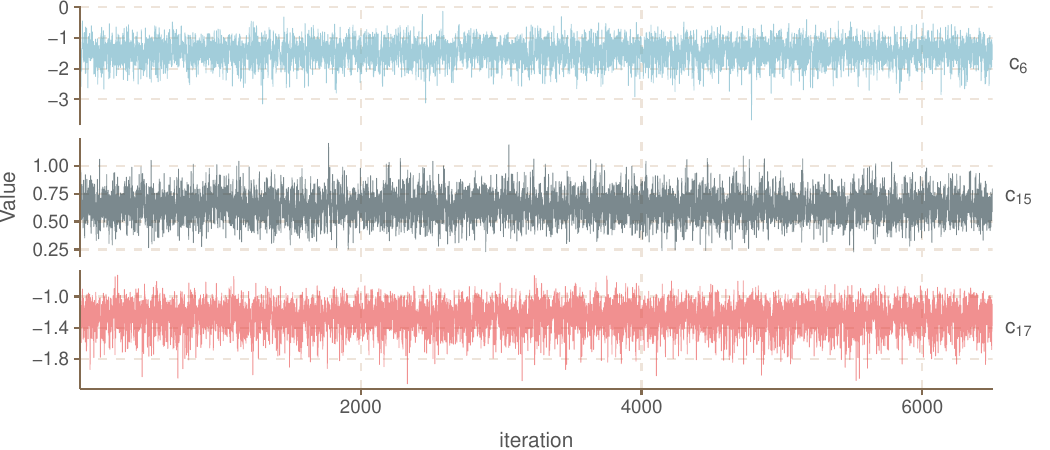}
  \caption{Traceplots of difficulty parameters: only 3 out of 18 were randomly selected to be shown.}
  \label{spifa_fig:trace_c}
\end{figure}

\begin{figure}[H]
  \centering
  \includegraphics[width=\linewidth]{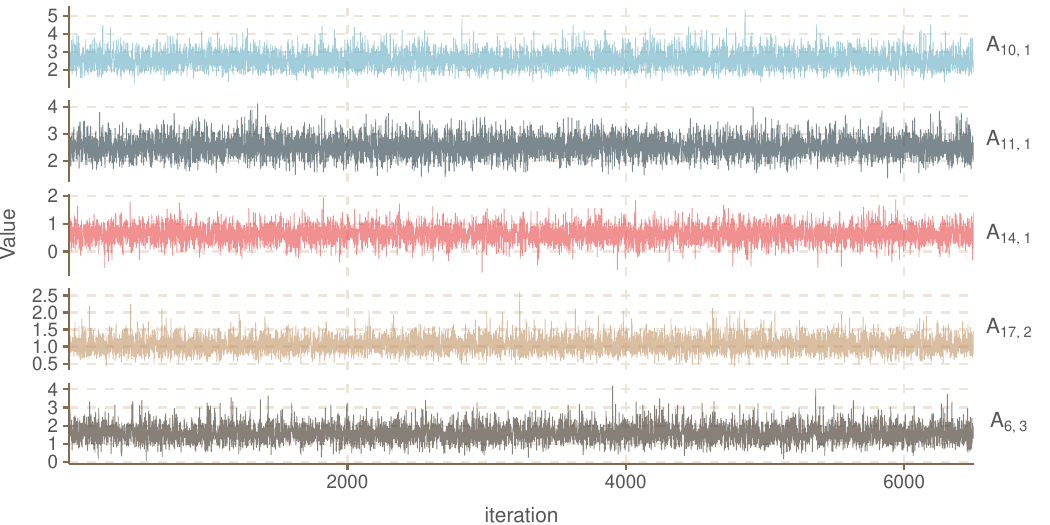}
  \caption{Traceplots of discrimination parameters: only 5 out of 22 were randomly selected to be shown.}
  \label{spifa_fig:trace_a}
\end{figure}

\begin{figure}[H]
  \centering
  \includegraphics[width=\linewidth]{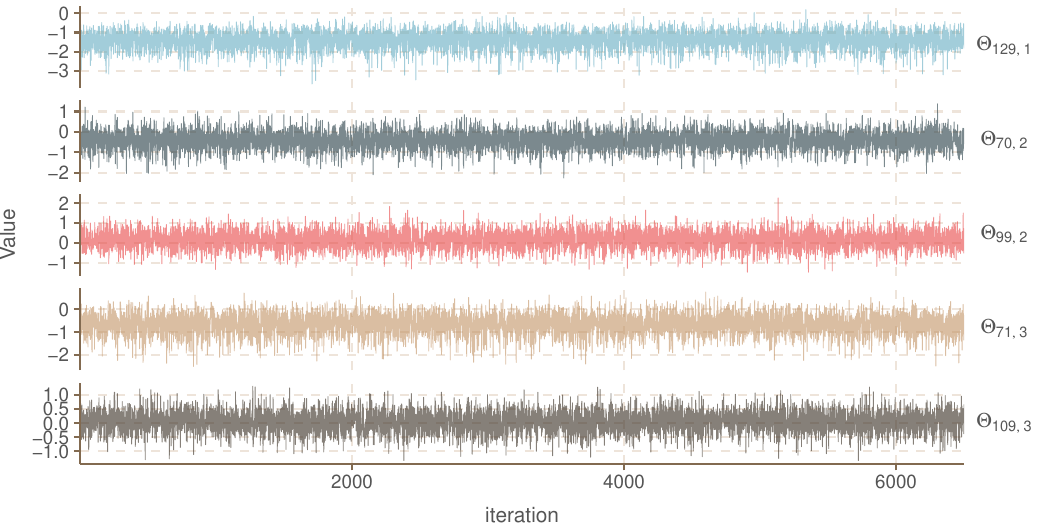}
  \caption{Traceplots of discrimination parameters: only 5 out of 600 were randomly selected to be shown.}
  \label{spifa_fig:trace_theta}
\end{figure}

\begin{figure}[H]
  \centering
  \includegraphics[width=\linewidth]{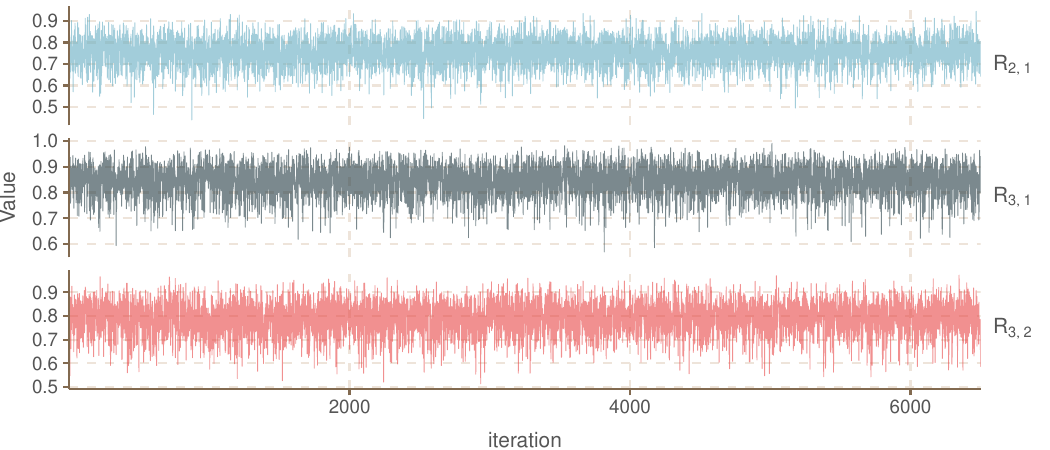}
  \caption{Traceplots of correlation parameters.}
  \label{spifa_fig:trace_corr}
\end{figure}

\begin{figure}[H]
  \centering
  \includegraphics[width=\linewidth]{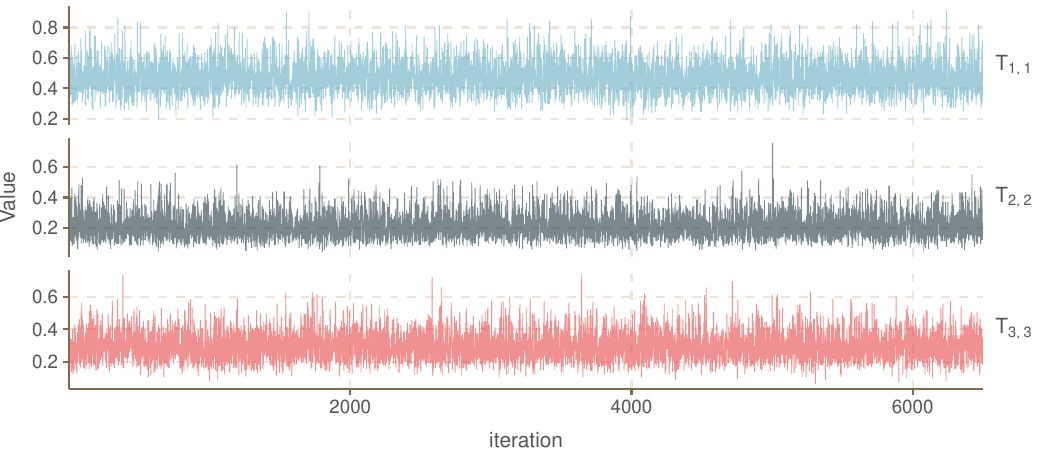}
  \caption{Traceplots of unrestricted standard deviations parameters for the multivariate Gaussian process.}
  \label{spifa_fig:trace_sd}
\end{figure}

\begin{figure}[H]
  \centering
  \includegraphics[width=\linewidth]{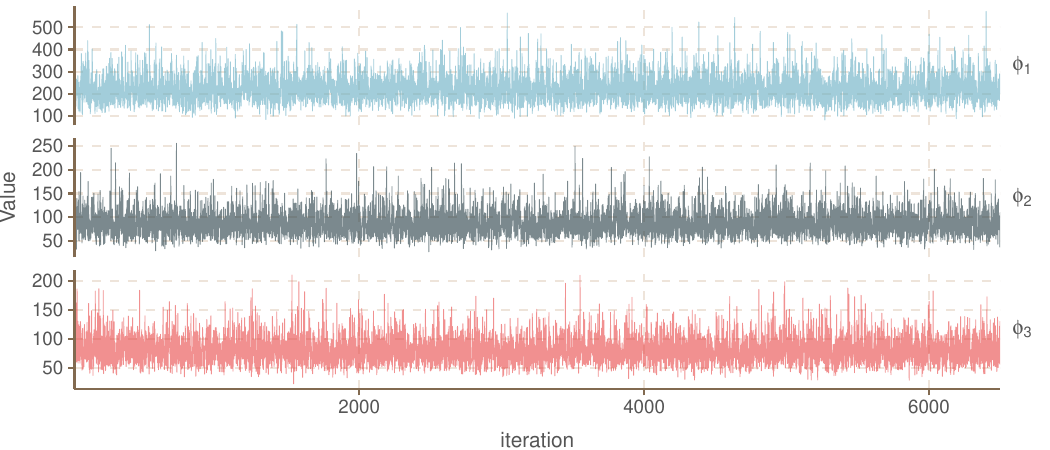}
  \caption{Traceplots of the scale parameters of the multivariate Gaussian process.}
  \label{spifa_fig:trace_phi}
\end{figure}

\subsection{Missing data comparison}
\label{spifa_app:missing}

\begin{figure}[H]
  \centering
  \includegraphics[width=0.95\textwidth]{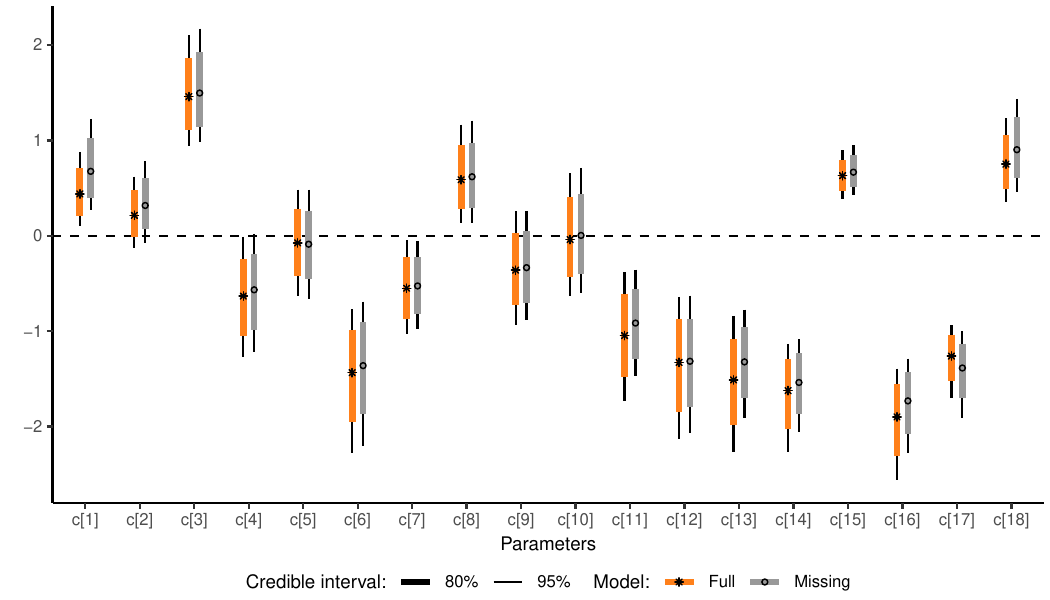}
  \caption{Credible intervals of the easiness parameters using all data (Full) and after
  removing missing profiles (Missing).}
  \label{spifa_fig:ci_easiness}
\end{figure}

\begin{figure}[H]
  \centering
  \includegraphics[width=0.95\textwidth]{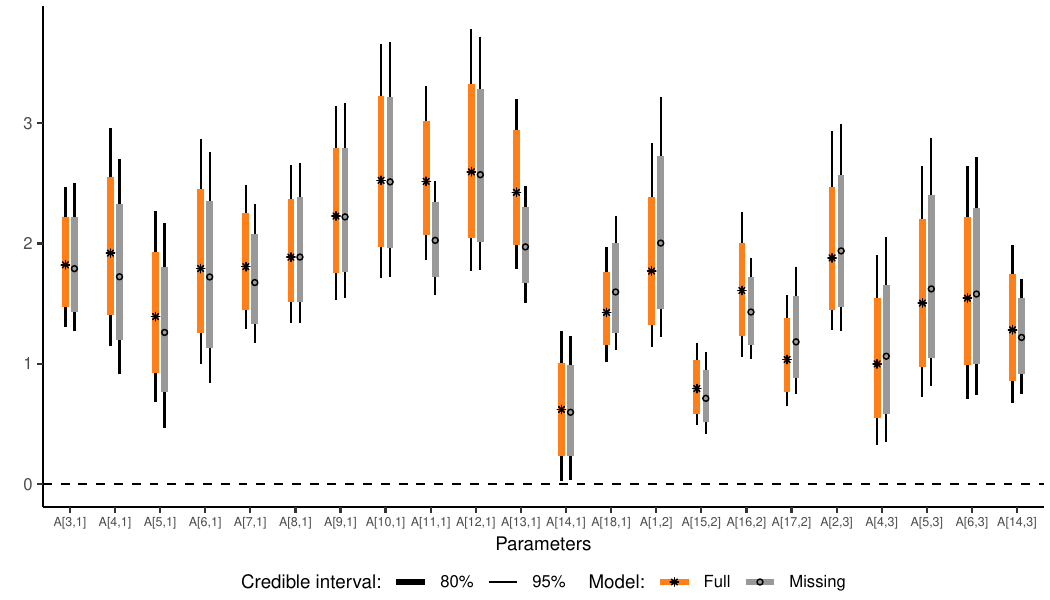}
  \caption{Credible intervals of the discrimination parameters $A[i,j]$ for item $i$ with
  respect to factor $j$ using all data (Full) and after removing missing profiles
  (Missing).}
  \label{spifa_fig:ci_discrimination}
\end{figure}

Figures \ref{spifa_fig:ci_easiness} and \ref{spifa_fig:ci_discrimination} compare the
estimated easiness and discrimination parameters, respectively, under the selected model
(SPIFA III) with those from an equivalent model that excludes profiles with missing items.
The results indicate no significant differences between the models, supporting the
assumption that the missing data mechanism is missing at random (MAR).

\subsection{Exceedance probabilities of food insecurity}
\label{spifa_app:excedances}

\begin{figure}[H]
  \centering
  \includegraphics[width=\linewidth]{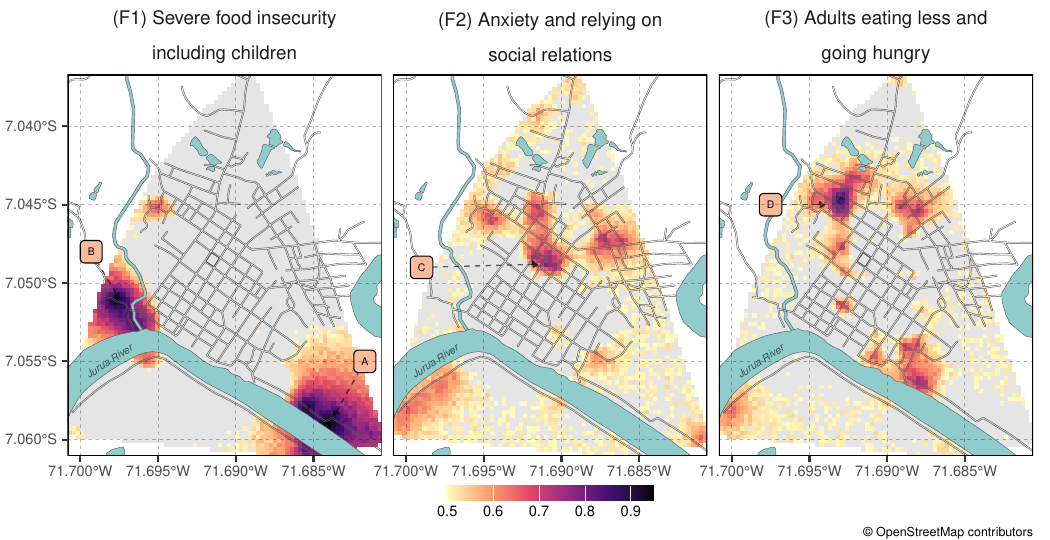}
  \caption{
    Probability of positive food insecurity dimensions ($> 0$). Only regions with
    probabilities $> 0.5$ are shown; labeled regions exceed $0.7$.
  }
  \label{spifa_fig:exceedance}
\end{figure}

\begin{figure}[H]
  \centering
  \includegraphics[width=\linewidth]{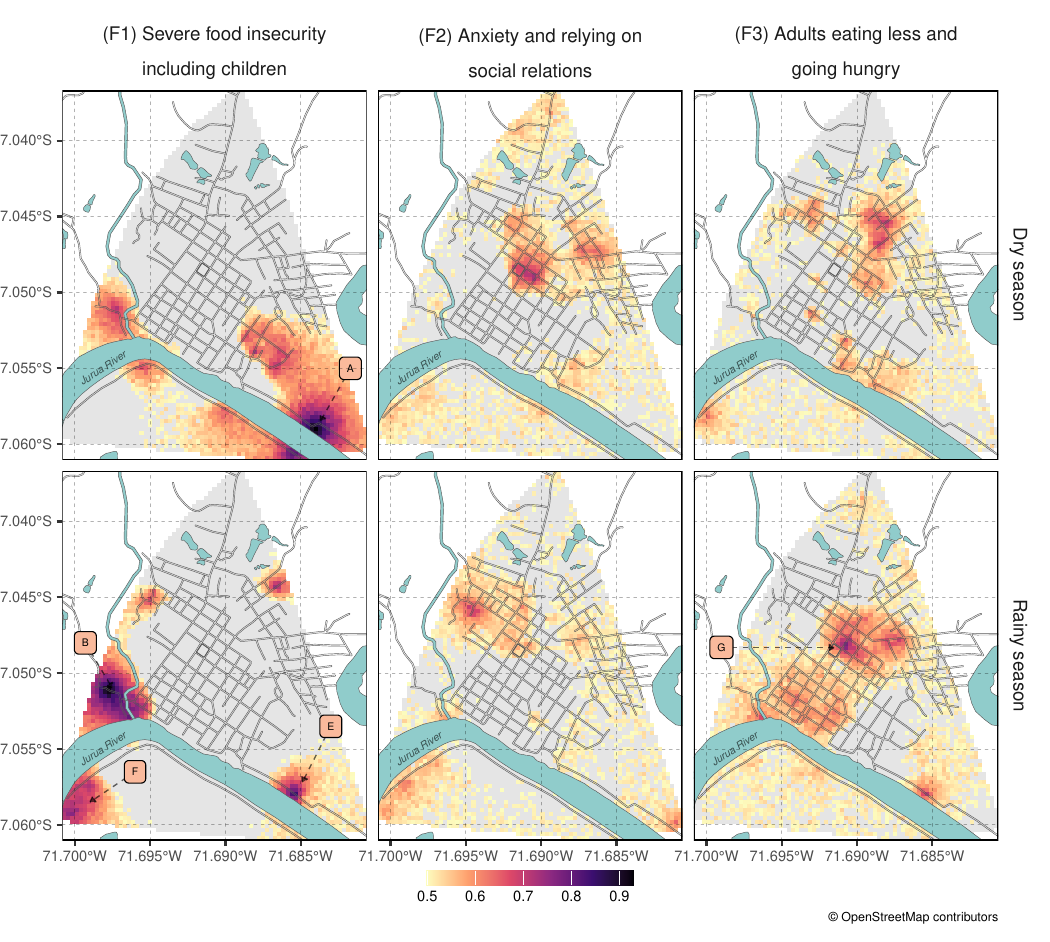}
  \caption{
    Probability of positive food insecurity dimensions ($> 0$) for the high-water season
    (Rainy) and low-water season (Dry). Only regions with probabilities $> 0.5$ are shown;
    labeled regions exceed $0.7$.
  }
  \label{spifa_fig:exceedance_season}
\end{figure}

Figures \ref{spifa_fig:exceedance} and \ref{spifa_fig:exceedance_season} display the
probabilities of food insecurity exceeding zero, using the full dataset and separated by
season, respectively. Regions with exceedance probabilities above 0.7 are identified as
high food insecurity zones.

\section{Extension to Mixed Outcome Types \label{sect:mixed_outcome}}

Some applications require jointly modeling different types of data, such as binary,
ordinal, categorical, and/or continuous \citep[e.g.,][]{mcparland2017clustering}.
In order to deal with binary, ordinal or continuous items, we can extend the spatial item factor analysis by considering $q_1$ ordinal items and $q_2$ continuous items. We do not need to differentiate another set of binary items given the they are simply ordinal items with two categories. The $q_1$ ordinal times can be modelled as spatial discrete-valued stochastic processes $\{\rv{Y}_j(s): s \in D\}$, where $D \subset \real^2$ and the random variable $\rv{Y}_j(s)$ can take values $\{0, 1, \dots, K_j-1\}$. Notice that $K_j$ represents the number of categories for ordinal item $j = 1, \dots, q_1$. We assume that the values of the $q_1$ discrete-valued stochastic processes are determined by an auxiliary real-valued stochastic processes $\{\rv{Z}^o_j(s): s \in D\}$ and thresholds $\ve{\gamma}_{j} = (\gamma_{j1}, \gamma_{j2}, \dots, \gamma_{j(K_j-1)})^\tr$ such as
\begin{align}
  \rv{Y}_{j}(s) = k \iff - \gamma_{jk} \leq \rv{Z}^o_{j}(s) < - \gamma_{j(k+1)}, ~\text{for}~ k = 0, 1, \dots, (K_j-1),
  \nonumber
\end{align}
where $\gamma_{j0} = - \infty$ and $\gamma_{j(K_j)} = \infty$. The $q_2$ continuous items can be modelled as real-valued stochastic processes $\{\rv{Z}^c_j(s): s \in D\}$ for $j = 1, \dots, q_2$. Then, we can defined the spatial random vector $\rve{Z}(s) = (\rv{Z}_1^o(s), \dots, \rv{Z}_{q_1}^o(s), \rv{Z}_1^c(s), \dots, \rv{Z}_{q_2}^c(s))^\tr$, a collection of the auxiliary random variables $\rv{Z}^o_j(s)$ associated to the ordinal items and the observable random variables $\rv{Z}^c_j(s)$ associated to the continuous items, and define the factor model at this level such as
\begin{align}
  \rv{Z}_{j}(s) & =
  c_j + \ve{a}_j^{*\intercal}\ve{\theta}(s) + \epsilon_{j}(s), ~\text{for}~ j = 1, \dots, q_1 + q_2 \nonumber
\end{align}
where, due to identifiability, $c_j = 0$ for $j=1, \dots, q_1$ and
where the $m$-dimensional latent factors are modelled including multivariate non-linear effects, $\ve{f}(x_j(s)): \real \to \real^m$,
\begin{align}
  \ve{\theta}(s) & = \sum_{i=1}^p\ve{f}(x_i(s))+ \mgp^*(s) + \ve{v}(s). \nonumber
\end{align}
Finally, to make the model identifiable, the error term is defined as
\begin{align}
  \epsilon_{j}(s) & \sim
  \left\lbrace
  \begin{array}[2]{cl}
    \df{N}(0, 1) & \text{for} ~ j = 1, \dots, q_1 \\
    \df{N}(0, \sigma_j) & \text{for} ~ j = q_1 + 1, \dots, q_1 + q_2 \\
  \end{array}
  \right..
  \nonumber
\end{align}

\end{appendices}

\end{document}